\documentclass{aastex6}

\begin{document}

\title{Evolution of molecular clouds in the superwind galaxy NGC 1808 probed by ALMA observations}

\author{Dragan Salak\altaffilmark{1}, Yuto Tomiyasu\altaffilmark{2}, Naomasa Nakai\altaffilmark{2,3}, Nario Kuno\altaffilmark{2,3}, Yusuke Miyamoto\altaffilmark{4}, and Hiroyuki Kaneko\altaffilmark{4}}

\altaffiltext{1}{School of Science and Technology, Kwansei Gakuin University, 2-1 Gakuen, Sanda, Hyogo 669-1337, Japan, d.salak@kwansei.ac.jp}
\altaffiltext{2}{Division of Physics, Faculty of Pure and Applied Sciences, University of Tsukuba, 1-1-1 Tennodai, Tsukuba, Ibaraki 305-8571, Japan}
\altaffiltext{3}{Center for Integrated Research in Fundamental Science and Technology,
University of Tsukuba, 1-1-1 Tennodai, Tsukuba, Ibaraki 305-8571, Japan}
\altaffiltext{4}{Nobeyama Radio Observatory, National Astronomical Observatory of Japan, 462-2 Nobeyama, Minamimaki, Minamisaku, Nagano 384-1305, Japan}

\begin{abstract}

ALMA imaging of the cold molecular medium in the nearby starburst galaxy NGC 1808 is presented. The observations reveal the distribution of molecular gas, traced by \(^{12}\)CO (1-0) and \(^{12}\)CO (3-2), and continuum (93 and 350 GHz) across the central 1 kpc starburst region at high resolution of \(\sim1\arcsec\). A molecular gas torus (radius \(\sim30\) pc) is discovered in the circumnuclear disk (CND; central 100 pc), with a high CO (3-2)/CO (1-0) ratio of \(\sim1\), surrounded by massive (\(10^6\)-\(10^7~M_\sun\)) clouds with high star formation efficiency (\(\mathrm{SFE}\sim10^{-8}\) yr\(^{-1}\)), molecular spiral arms, and a 500 pc pseudoring. The CND harbors a continuum core and molecular gas exhibiting peculiar motion. The new data confirm the line splitting along the minor galactic axis, interpreted as a nuclear gas outflow with average velocity \(\sim180\) km s\(^{-1}\), and show evidence of a velocity gradient of \(\sim+0.4\) km s\(^{-1}\) pc\(^{-1}\) along the axis. In addition, supershells expanding from the 500 pc ring with maximum velocities of \(\sim75\) km s\(^{-1}\) are revealed. The distribution and CO luminosities of molecular clouds in the central 1 kpc starburst region indicate an evolutionary sequence, from gas accretion onto the 500 pc ring from the large-scale bar, to enhanced star formation in the ring, and outflow as feedback.

\end{abstract}


\section{Introduction}\label{A}

In starburst galaxies, vigorous star formation inflicts feedback on the interstellar medium (ISM) in the form of stellar winds and supernova explosions that drive galactic superwinds (e.g., \citealt{CC85,MQT05,MMT11,VCB05,NS09,Sha11,Cre13,Roy13}). On the other hand, in active galactic nuclei (AGN), gas accretion onto the central supermassive black hole (SMBH) triggers feedback as intense radiation from the accretion disk and jets of hot gas (e.g., \citealt{IF15}). The kiloparsec-scale, multiphase outflows of gas and dust, frequently observed in starbursts and AGN, are thought to regulate star formation and SMBH feeding in galaxies (e.g., \citealt{Mar05,SB10,Sek16,Zsc16}). In particular, galactic winds can suppress star formation by removing cold molecular gas from galactic disks \citep{Bol13a,Cic14,Gea14,Alat15}, ignite star formation elsewhere by large-scale shocks (e.g., \citealt{Lac17}), transport heavy elements into galaxy halos, and cut the supply of cold gas toward accretion disks around SMBHs in AGN (e.g., \citealt{ZK12}). Recent observations also indicate that star formation may be possible even inside powerful outflows \citep{Mai17}.

Revealing the distribution and kinematics of the cold interstellar gas at the base of a starburst-driven outflow (in galactic central regions), where molecular clouds are thought to be entrained, is one of the key observational challenges. Sensitive observations at high resolution allow us to distinguish individual star-forming regions and investigate their link to the winds (e.g., \citealt{Wei99,Mat00,Bol13a,Sal16}). Such observations can also constrain the outflow energetics by yielding the mass of molecular gas and the outflow velocity, which determines whether the gas can escape from the host galaxy (e.g., \citealt{Bol13a,Sal16}). However, the process of entrainment of molecular clouds into outflows and their evolution remains poorly understood due to the difficulty of acquiring high-resolution images of molecular gas in starburst-driven winds.

On the other hand, starburst activity requires gas supply -- accumulation of large molecular gas mass in the galactic central region. Gas transport toward the central regions of galaxies is thought to be accomplished via gravitational torques induced by a bar, spiral arms, galaxy mergers, etc. (e.g., \citealt{Sch84,WH92,BC96,KK15,Kru17}). While there has been significant progress in understanding the gas dynamics in numerical simulations, high-resolution velocity field data of molecular gas, that can probe the gas kinematics in the central regions of barred superwind galaxies, are still sparse.

In this work, we have selected the nearby starburst galaxy NGC 1808 as a case-study of a starburst-driven superwind and conducted comprehensive multi-line imaging using the Atacama Large Millimeter/submillimeter Array (ALMA) in cycles 1 and 2, supplemented by high-resolution radio continuum observations using the Very Large Array (VLA). In this second paper of a series about NGC 1808 based on ALMA observations, we focus on the distribution and kinematics of molecular clouds in the central 1 kpc starburst region traced by the CO (1-0) and CO (3-2) lines at high resolution (\(\sim1\arcsec\) or \(\sim50\) pc), derive their masses, star-forming properties, and investigate the structure and kinematics of the molecular wind. As described below, the galaxy is an excellent target to address these issues because of proximity (\(10.8~\mathrm{Mpc}\); \citealt{Tul88}) and large abundance of molecular gas.

\begin{figure}
\epsscale{1.15}
\plotone{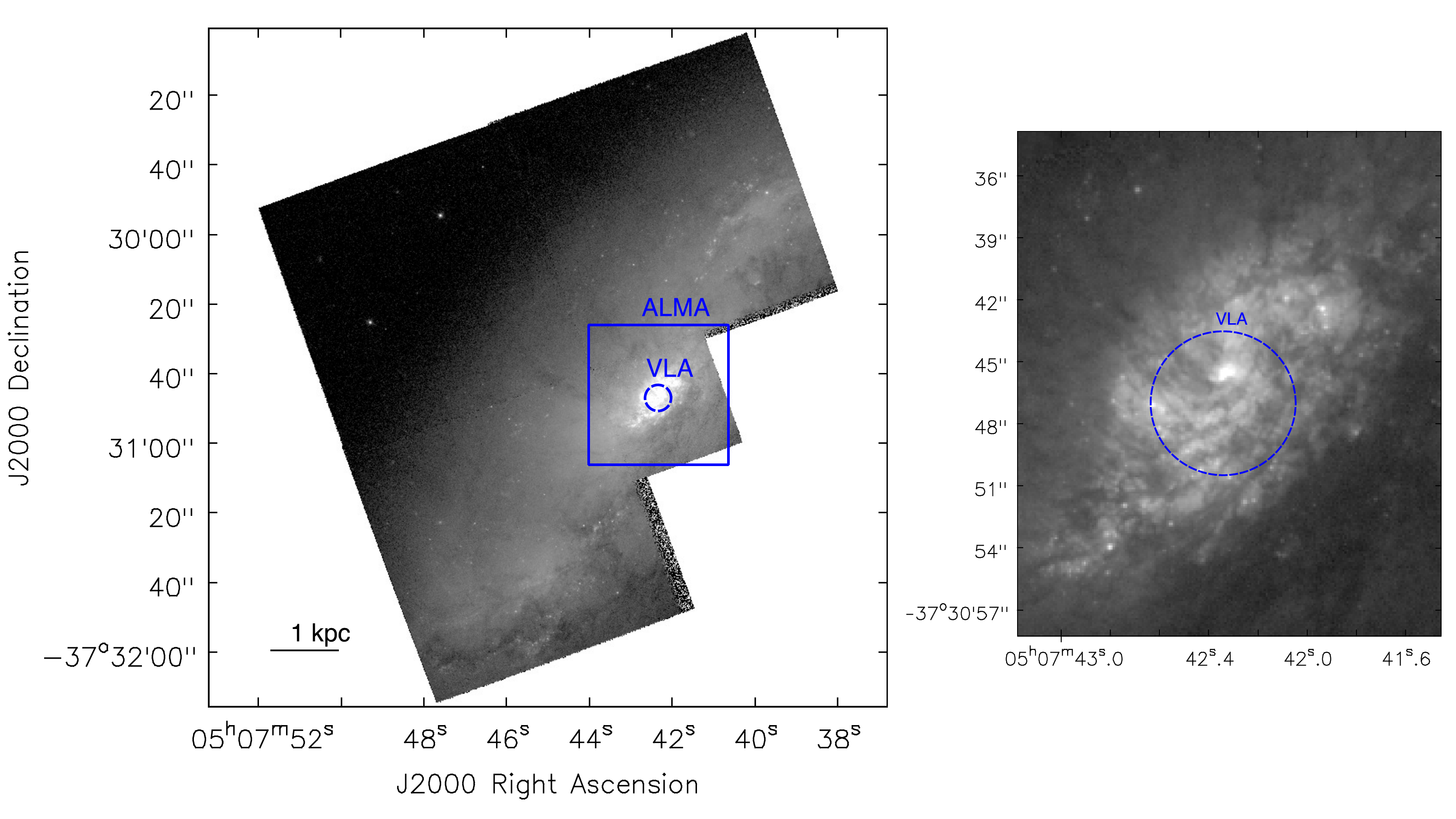}
\caption{Optical R-band image of NGC 1808 (acquired from the Hubble Legacy Archive). The rectangle and circle show the image size of ALMA cycle 2 observations (band 7) and the field of view of VLA K-band (21 GHz) observations, respectively.\label{fig:n1808}}
\end{figure}

NGC 1808 is a barred spiral galaxy with a starburst in its central \(r\sim500~\mathrm{pc}\) region (Figure \ref{fig:n1808}, Table \ref{tab1}). Its peculiar optical morphology with polar dust lanes, conspicuous in the color index images in \cite{VV85}, \cite{Phi93}, and the \emph{Carnegie-Irvine Galaxy Survey} \citep{Ho11}, resembles the ``explosive'' galactic nucleus observed in the nearby starburst galaxy M82 \citep{BB68}. The galaxy's moderate inclination (\(57\degr\)) allows us to get a clear picture of the distribution and kinematics of stellar and ISM components over the galactic disk. Previous studies have revealed numerous supernova remnants and \ion{H}{2} regions in the central \(r\sim500~\mathrm{pc}\) \citep{Con87,Sai90,FBW92,KSG94,Kot96}, known as the nuclear ``hot spots'' \citep{Mor58,SP65}. Infrared observations have revealed a nuclear bar and star clusters \citep{TSE96,TG05,Gal05,GA08,Bus17}. The nucleus (central \(r<100~\mathrm{pc}\)) shows activity that can be attributed to a starburst possibly coexisting with a weak (low-luminosity) AGN or ultra-luminous X-ray sources (ULXs) generating hard X-ray emission \citep{VV85,Jun95,Awa96,Lau00,JB05,HA07}. In the central kiloparsec, molecular gas is distributed in a circumnuclear disk (CND) within \(r<100~\mathrm{pc}\), a disk dominated by nuclear spiral arms within \(r<400~\mathrm{pc}\), and a resonant pseudoring at the ILR (\(r\sim500~\mathrm{pc}\)) \citep{Sal16}.  Noncircular motions were also found in the starburst region indicating inflow and outflow of molecular gas.

The paper is organized as follows. We begin with a description of the observations of radio continuum and molecular spectral lines carried out with the VLA and ALMA telescopes (section \ref{B}). In section \ref{C}, we present the main results, including the distribution and intensity of the continuum and CO lines, and report the discovery of a molecular gas torus in the galactic center. A discussion in sections \ref{D} and \ref{F} is focused on star formation properties and molecular gas outflows, respectively. Based on gas distribution, CO line intensity ratios, and kinematics, we propose an evolutionary sequence of molecular clouds in the central starburst region, driven by dynamics, star formation activity and its feedback (section \ref{E}). The article is concluded with a summary in section \ref{G}.

The velocities in this paper are presented with respect to the local standard of rest (LSR) in radio definition.

\begin{table}
\begin{center}
\caption{Basic Parameters of NGC 1808}\label{tab1}
\begin{tabular}{llc}
\tableline\tableline
Parameter & Value & Reference \\
\hline
Morphological type & (R)SAB(s)a & (1) \\
Right ascension (J2000.0) & \(05^\mathrm{h}07^\mathrm{m}42\fs331\pm0\fs003\) & (2) \\
Declination (J2000.0) & \(-37\degr30\arcmin45\farcs88\pm0\farcs05\) & (2) \\
Distance & \(10.8~\mathrm{Mpc}\) (\(1\arcsec=52\) pc) & (3) \\
Systemic velocity (LSR) & \(998~\mathrm{km~s}^{-1}\) (central \(200~\mathrm{pc}\) region) & (4) \\
Position angle & \(324\degr\) & (4) \\
Inclination & \(57\degr\) & (5) \\
Activity & \ion{H}{2}, Seyfert 2 & (6) \\
\tableline
\end{tabular}
\end{center}
\tablecomments{References: (1) \cite{deV91}, (2) this work (93 GHz continuum), (3) \cite{Tul88}, (4) \cite{Sal16}, (5) \cite{Rei82}, (6) NED classification.}
\end{table}

\section{Observations}\label{B}

\subsection{VLA Observations}\label{Ba}

The observations with the Very Large Array (VLA) were conducted in 2012, during the Expanded VLA project (EVLA) of upgrading the capabilities of the interferometer, with 27 antennas in configuration BnA. Since the target galaxy is at declination \(\delta\approx-38\degr\), this configuration yields the highest angular resolution with well-sampled spatial frequencies. The imaging was carried out with a single pointing toward the phase center at \((\alpha,\delta)_\mathrm{J2000}=(05^\mathrm{h}07^\mathrm{m}42\fs34,-37\degr30\arcmin46\farcs98)\), corresponding to the galactic center from the \emph{Two Micron All Sky Survey} (2MASS).

The spectral configuration of the K band (frequency 21 GHz) was arranged to cover two line-free bands centered at \(20.7\) and \(21.7~\mathrm{GHz}\), each with a width of \(1.024~\mathrm{GHz}\), yielding a total synthesized continuum bandwidth of \(2.02404~\mathrm{GHz}\) with the center frequency at \(21.1993~\mathrm{GHz}\). In the Ka band (32 GHz), the bands were centered at \(31.52\) and \(32.52~\mathrm{GHz}\). The total on-source time was 37 minutes in K band and 100 minutes in Ka band.

Data reduction was done using the Common Astronomy Software Applications (CASA) program \citep{McM07}. The data were first reduced via the EVLA pipeline, then split and interactively imaged (algorithm CLEAN) in multi-frequency synthesis mode with Briggs weighting (robustness parameter set to 0.5). The resulting synthesized beam was \(0\farcs46\times0\farcs23\) (K) and \(0\farcs23\times0\farcs18\) (Ka) (full width at half-maximum; FWHM). This is equivalent to \(24~\mathrm{pc}\times12~\mathrm{pc}\) and \(12~\mathrm{pc}\times9~\mathrm{pc}\), respectively. The Ka band image is the highest angular resolution image of the nucleus of NGC 1808. The largest angular scale\footnote{The maximum size of the structure that can be sampled by the interferometer without short-spacing correction.} visible to the array was \(\sim7\arcsec\) (K; dotted circle in Figure \ref{fig:n1808}) and \(\sim4\arcsec\) (Ka) fully covering the region of the circumnuclear disk (diameter \(\sim3\arcsec\), \citealt{Sal16}). Both images are corrected for the primary beam attenuation. A summary of the observations is given in Table \ref{tab2}.

\subsection{ALMA Observations}\label{Bb}

\subsubsection{Band 3}\label{Bba}

The observations in band 3 were performed toward the galactic center (single pointing) with the 12 m array during 2014 and 2015 as part of our cycle 2 project. The number of antennas increased from 34 to 41 during the cycle. Band 3 observations were conducted using two configurations of the 12 m array (extended and compact) yielding a good \((u,v)\) coverage. As a result, the largest angular scale, given by \(\sim0.6\lambda/L_\mathrm{min}\) where \(L_\mathrm{min}\simeq15~\mathrm{m}\) is the shortest projected baseline, is \(\sim26\arcsec\), which is larger than the central starburst region (\(\sim20\arcsec\)). The total on-source time was about 24 minutes. In this paper we focus on the results of continuum observations, whereas band 3 spectral lines observed in cycle 2 will be presented elsewhere. Data were acquired from four sub-bands centered at the rest frequencies of 86.800, 88.716, 97.981, and 100.750 GHz, with total widths of 1.875, 1.875, 0.235, and 2.000 GHz, respectively, giving a representative frequency of 93 GHz.

The data were calibrated, deconvolved, and corrected for the primary beam attenuation using the CASA package. The spectral lines were flagged prior to continuum imaging. The imaging was done interactively with Briggs weighting (robustness parameter set to 0.5) to achieve high angular resolution. The final data cube has an rms sensitivity of \(25~\mathrm{\mu Jy~beam}^{-1}\) and the synthesized beam of \(1\farcs17\times0\farcs77\) (FWHM). The images, corrected for the primary beam attenuation, show no indication of sidelobes suggesting that missing flux is likely negligible.

In this paper, we also use CO (1-0) data from cycle 1, that have been corrected for the missing flux. The data acquired by the 12 m array only were previously presented in \cite{Sal16}. The 12 m array data were first CLEANed together with Atacama Compact Array (ACA) data, and then corrected for the missing flux relative to the Total Power (TP) single-dish data cube using the CASA task ``feather''. The sensitivity and angular resolution of the final CO (1-0) data cube are \(5~\mathrm{mJy~beam^{-1}}\) (channel width \(\Delta v=10\) km s\(^{-1}\)) and \(2\farcs67\times1\farcs48\) (\(\sim100~\mathrm{pc}\)), respectively. A more detailed presentation of the CO (1-0) data (corrected for missing short baselines) will be given in a separate paper.

\subsubsection{Band 7}\label{Bbb}

Band 7 observations took place during 2014 and 2015 with 38 antennas of the 12 m array, 9-10 antennas of the ACA, and 3 antennas of the TP. The mosaic image (\(40\arcsec\times40\arcsec\) equivalent to \(2.08~\mathrm{kpc}\times2.08~\mathrm{kpc}\)) was covered with 27 pointings of the 12 m array and 11 pointings of ACA; the field was mapped by the TP antennas in single-dish observations. The target molecular line was CO (\(J=3-2\)) at the rest frequency of 345.79599 GHz, whereas all four available sub-bands, each with a width of 1.9-2.0 GHz, were used to image the continuum. The sub-bands were centered at the rest frequencies of 344.000, 345.796, 356.734, and 358.000 GHz. In the same band, the HCO\(^{+}\) (4-3) line was observed simultaneously with CO, and the results will be presented together with band 3 spectral lines in a separate paper. The CO and HCO\(^{+}\) lines were flagged prior to the continuum imaging. The imaging was done interactively in mosaic mode with natural weighting to maximize sensitivity, and the line data acquired by the three arrays were combined into a single data cube using the CASA task ``feather''. Before the combining process, the TP data were thoroughly flagged to remove the spectra with poor baselines. To check the total flux, we compared the combined ALMA data to the single-dish ASTE data from \cite{Sal14}. The discrepancy is a few percents, which is within the uncertainty of the intensity calibration of single-dish data.

The final CO (3-2) data cube has a sensitivity of \(8~\mathrm{mJy~beam}^{-1}\) (velocity resolution \(\Delta v=5~\mathrm{km~s}^{-1}\)), and FWHM of the synthesized beam is \(1\farcs04\times0\farcs57\) (\(54~\mathrm{pc}\times30~\mathrm{pc}\)). The continuum image at 350 GHz was produced by combining the data acquired by the 12 m array and ACA, yielding a synthesized beam of \(1\farcs04\times0\farcs56\) (\(54~\mathrm{pc}\times29~\mathrm{pc}\)). Natural weighting was applied and the resulting sensitivity is \(0.35~\mathrm{mJy~beam}^{-1}\) over a total bandwidth of \(7~\mathrm{GHz}\). All images were corrected for the primary beam attenuation.

The details about the observations in bands 3 and 7 in cycle 2 are summarized in Table \ref{tab2}.

\begin{table}
\begin{center}
\caption{Observational Summary}\label{tab2}
\begin{tabular}{lllll}
\tableline\tableline
Parameter & VLA (K) & VLA (Ka) & ALMA (B3) & ALMA (B7) \\
\hline
Representative frequency [GHz] & 21 & 32 & 93 & 350 \\
Observation date & 2012 Sep 8 & 2012 Sep 8, 13 & 2014 Aug 27, Dec 7  & 2014 Jun 14, Dec 13 \\
& & & 2015 Apr 21 & 2015 May 1, Jun 28 \\
Number of antennas & 27 & 27 & 34-41 & \(38+\mathrm{ACA}10+\mathrm{TP}3\) \\
Bandpass calibrator & 3C147 & 3C147 & J0519-4546 & J0522-3627 \\
Flux calibrator & 3C147 & 3C147 & J0522-364 & J0522-364 \\
Phase calibrator & J0424-3756 & J0424-3756 & J0522-3627 & J0522-3627 \\
Number of pointings & 1 & 1 & 1 & \(27+11\) \\
Mosaic size & \nodata & \nodata & \nodata & \(40\arcsec\times40\arcsec\) \\
Synthesized beam\tablenotemark{a} (\(\theta_\mathrm{maj}\times\theta_\mathrm{min}\)) & \(0\farcs46\times0\farcs23\) & \(0\farcs23\times0\farcs18\) & \(1\farcs17\times0\farcs77\) & \(1\farcs04\times0\farcs56\) \\
Beam position angle & \(26\degr\) & \(-6\degr\) & \(75\degr\) & \(84\degr\) \\
Total time on source [min.] & 47 & 100 & 24 & \(9+27+73\) \\
Sensitivity (\(1\sigma\), continuum) & \(0.09~\mathrm{mJy~beam}^{-1}\) & \(0.06~\mathrm{mJy~beam}^{-1}\) & \(0.025~\mathrm{mJy~beam}^{-1}\) & \(0.35~\mathrm{mJy~beam}^{-1}\) \\
Sensitivity (\(1\sigma\), line) & \nodata & \nodata & \(1~\mathrm{mJy~beam}^{-1}\) & \(8~\mathrm{mJy~beam}^{-1}\) \\
\tableline
\end{tabular}
\end{center}
\tablenotemark{a}{Refers to the continuum images. \(\theta_\mathrm{maj}\) and \(\theta_\mathrm{min}\) are the major and minor axes of the beam FWHM.}
\end{table}

\section{Results}\label{C}

\subsection{Continuum emission}\label{Ca}

\begin{figure}
\plotone{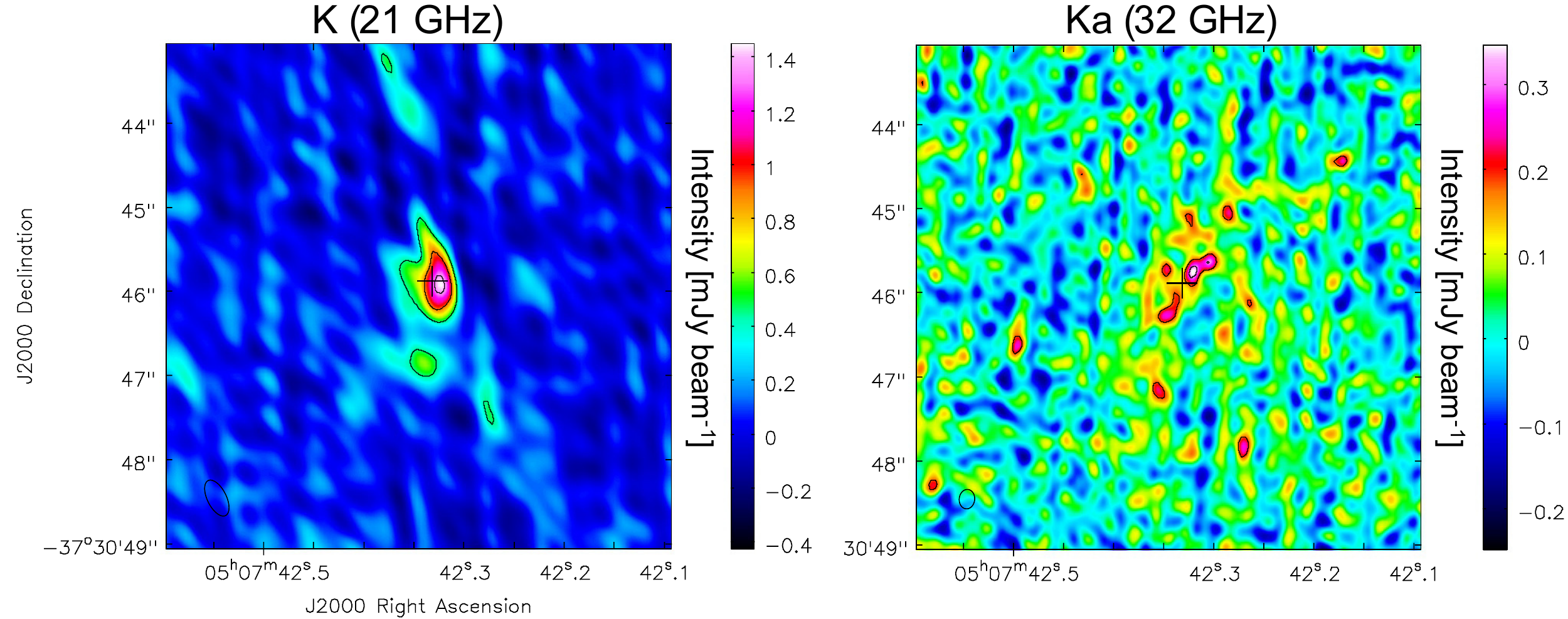}
\caption{K (left) and Ka (right) images toward the galactic center. The colorbar is displayed up to maximum values, where \(1~\sigma_\mathrm{K}=9\times10^{-5}~\mathrm{Jy~beam}^{-1}\), \(\mathcal{S}_\mathrm{K,max}=1.45\times10^{-3}~\mathrm{Jy~beam}^{-1}\), \(1~\sigma_\mathrm{Ka}=6\times10^{-5}~\mathrm{Jy~beam}^{-1}\), and \(\mathcal{S}_\mathrm{Ka,max}=3.5\times10^{-4}~\mathrm{Jy~beam}^{-1}\). The contours are plotted at \((5,10, 15)\times1~\sigma_\mathrm{K}\) and \((3,5)\times1~\sigma_\mathrm{Ka}\). The plus sign marks the 3 mm continuum peak (Table \ref{tab1}). The synthesized beam is shown at the bottom left corner. For the adopted galaxy distance, \(1\arcsec=52~\mathrm{pc}\).\label{fig:cont1}}
\end{figure}

The results of continuum observations in the VLA bands K (21 GHz) and Ka (32 GHz) are shown in Figure \ref{fig:cont1}. Emission was detected at \(>5~\sigma\) toward the galactic center in both bands. The images show a luminous core in the central \(1\arcsec\), composed of a marginally resolved source. The peak positions of the two continuum images, derived from 2D Gaussian fitting within a circular region of diameter \(0\farcs5\) centered at the brightest pixel, are \((\alpha,\delta)_\mathrm{21GHz}=(05^\mathrm{h}07^\mathrm{m}42\fs3266\pm0\fs0007,
-37\degr30\arcmin45\farcs928\pm0\farcs022)\) at 21 GHz, and \((\alpha,\delta)_\mathrm{32GHz}=(05^\mathrm{h}07^\mathrm{m}42\fs3146\pm0\fs0060,
-37\degr30\arcmin45\farcs704\pm0\farcs047)\) at 31 GHz. The peaks are separated by \(\simeq0\farcs2\) (10 pc), which is comparable to the beam size of the higher-resolution image.

\begin{figure}
\plotone{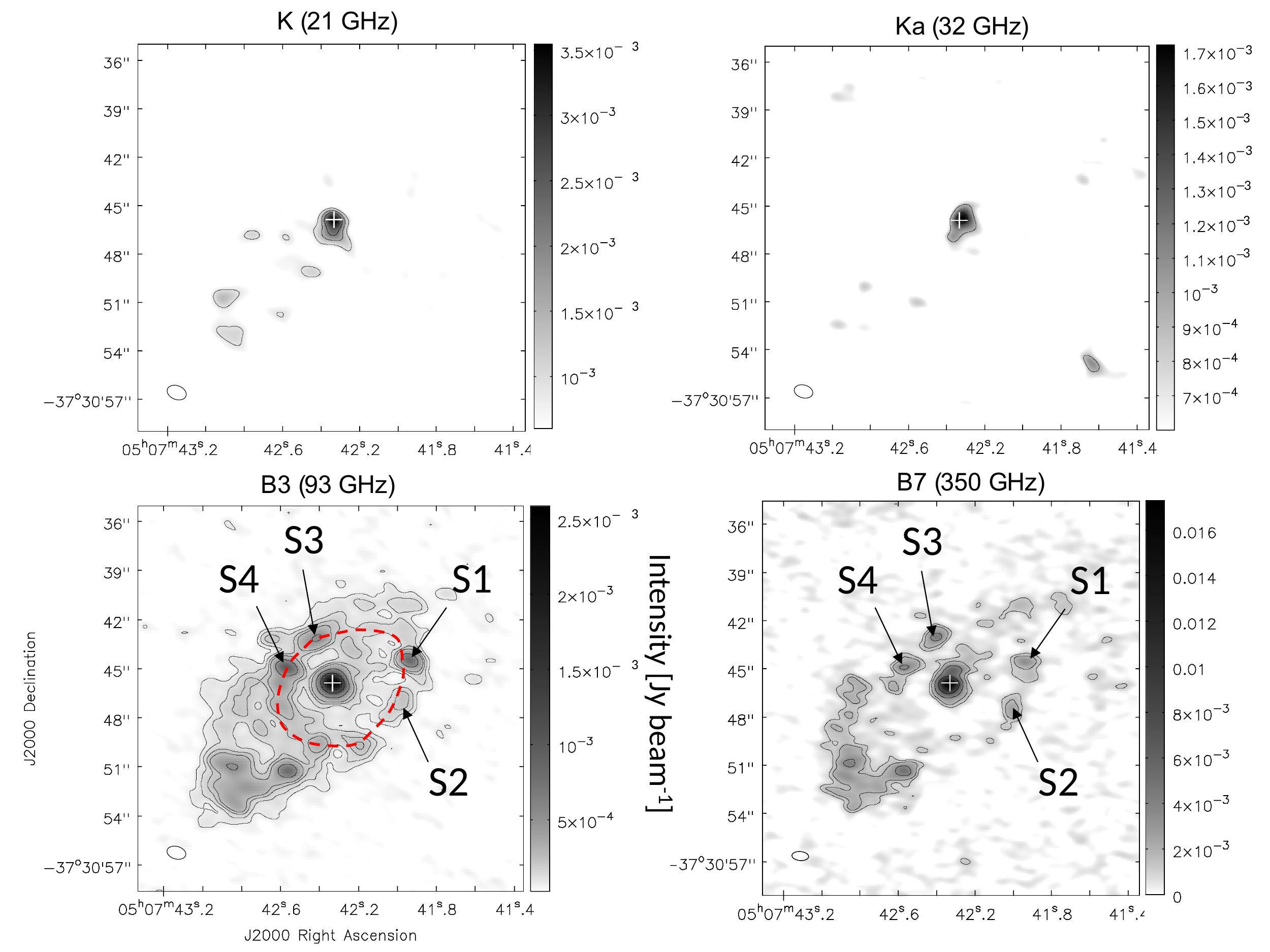}
\caption{Intensity images from all bands smoothed to the same angular resolution (band 3) to determine the total flux density and the spectral energy distribution in the nucleus. The contours are plotted as follows: K) \((3, 5, 7)\times0.3~\mathrm{mJy~beam}^{-1}(1~\sigma)\), Ka) \((3, 5)\times0.3~\mathrm{mJy~beam}^{-1}(1~\sigma)\), B3) \((3, 5, 7.5, 10, 20)\times0.025~\mathrm{mJy~beam}^{-1}(1~\sigma)\), and B7) \((3, 5, 10, 20)\times0.35~\mathrm{mJy~beam}^{-1}(1~\sigma)\) The maximum value of the B7 image at original resolution is \(17~\mathrm{mJy~beam}^{-1}\). The red dashed circle in panel B3 marks the star-forming ring of \ion{H}{2} regions harboring compact sources S1-S4.\label{fig:cont2}}
\end{figure}

The continuum images of ALMA bands 3 (93 GHz) and 7 (350 GHz), shown in Figure \ref{fig:cont2} together with the VLA images smoothed to the same angular resolution, reveal extended structure dominated by the core and a number of discrete sources within the central \(15\arcsec\). The structure of the 93 GHz continuum distribution is similar to that at lower frequencies (5 and 8 GHz) found in previous observations confirming the presence of radio hot spots \citep{Sai90,Col94}. The total flux densities and peak fluxes of the core are listed in Table \ref{tab:cont_flux}. The coordinates of the 93 GHz continuum peak, derived from 2D Gaussian fitting and adopted as the Galactic center position, are given in Table \ref{tab1}.

Figure \ref{fig:cont2} (panels B3 and B7) also shows a number of discrete 93 GHz and 350 GHz sources (S1-S4 ``hot spots'') distributed in a star-forming ring with a radius of \(\sim4\arcsec\) (200 pc; marked by a red dashed line) around the galactic center. The continuum emission from the sources was previously detected in a lower-resolution image in \cite{Sal16} and at lower frequencies by \cite{Sai90} and \cite{Col94}. The star-forming ring was studied by \cite{Gal05} and \cite{Bus17}, who used infrared data to derive the age of young star clusters associated with compact sources along the ring (5-8 Myr). The detection of discrete sources S1-S4 in the 93 GHz continuum is consistent with the picture that the objects are \ion{H}{2} regions emitting thermal free-free radiation produced in ionized gas. The sources also coincide with massive molecular clouds (central molecular clouds; CMCs) as will be shown in section 3.2.

\begin{table}
\begin{center}
\caption{Continuum Emission Flux in the Core}\label{tab:cont_flux}
\begin{tabular}{lrrrr}
\tableline\tableline
Parameter & K (21 GHz) & Ka (32 GHz) & B3 (93 GHz) & B7 (350 GHz) \\
\hline
Total flux density \(S_\nu~[\mathrm{mJy}]\) & \(6.79\pm0.72\) & \(6.19\pm0.58\) & \(4.19\pm0.23\) & \(39.8\pm2.5\) \\
Maximum intensity \(\mathcal{S}_\mathrm{max}~[\mathrm{mJy~beam}^{-1}]\) & \(3.41\pm0.26\) & \(1.62\pm0.12\) & \(2.45\pm0.09\) & \(19.21\pm0.86\) \\
\tableline
\end{tabular}
\end{center}
\tablecomments{The flux densities and peak intensities were calculated from two-dimensional Gaussian fitting over a circular region (diameter \(2\arcsec\)) after smoothing all images to the angular resolution of the B3 image. The region center is identical to the galactic center in Table \ref{tab1}. The maximum values of images at original resolution are given in Figures \ref{fig:cont1} and \ref{fig:cont2}.}
\end{table}

\subsubsection{Spectral index}\label{Caa}

Figure \ref{fig:cont3}(a) shows the spectral index produced from 93 GHz and 350 GHz continuum images in the central region, defined as

\begin{equation}
\alpha_\nu\equiv\frac{\log(S_\mathrm{u}/S_\mathrm{l})}{\log(\nu_\mathrm{u}/\nu_\mathrm{l})},
\end{equation}
where \(\nu_\mathrm{l}=93~\mathrm{GHz}\) and \(\nu_\mathrm{u}=350~\mathrm{GHz}\), and \(S_\mathrm{l}\) and \(S_\mathrm{u}\) are the corresponding flux densities. The index can be used as a diagnostic tool to probe the emission mechanism (synchrotron, free-free, and thermal radiation from dust grains) and/or the variation of dust temperature within molecular clouds. Typical values are \(\alpha_\mathrm{s}\simeq-0.9\) (synchrotron), \(\alpha_\mathrm{ff}\simeq-0.1\) (free-free), and  \(\alpha_\mathrm{d}\simeq3\mathrm{-}4\) (dust). The dust emission can be represented by a modified Planck function, \(\nu^\beta B_\nu(T)\), where \(\beta\) is the emissivity index (0 for blackbody and between 1 and 2 for most grain materials). In the Rayleigh-Jeans regime of the dust emission spectrum, the flux density becomes \(S_\nu\propto\nu^{2+\beta}\propto\nu^\alpha\), hence we expect \(3\leq\alpha\leq4\). Note that the structure in Figure \ref{fig:cont3}(a) is well-resolved and exhibits a significant variation within \(0.5\lesssim\alpha\lesssim2.5\). The image shows that the spectral index is \(\sim1.5\) in the core surrounded by a structure of enhanced values of \(\sim2\) in the central 100 pc (\(2\arcsec\)); \(\alpha\) is also enhanced in some outer regions.

\subsubsection{Spectral energy distribution of the core}\label{Cab}

\begin{figure}
\epsscale{1}
\plotone{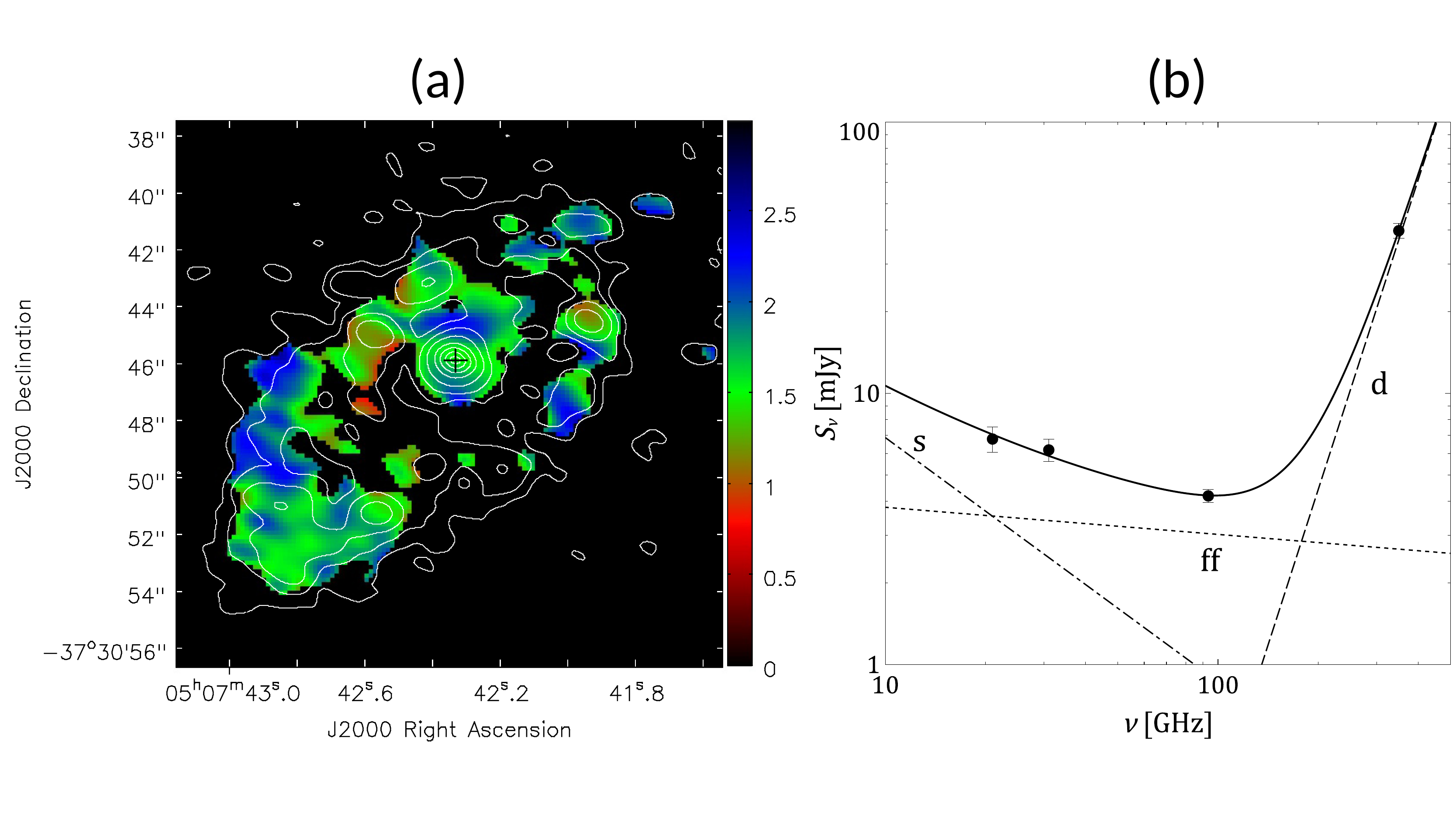}
\caption{(a) Spectral index between 93 and 350 GHz. The image was produced after adjusting the 93 and 350 GHz images to the same angular resolution and clipping at \(3~\sigma\). The black plus sign marks the 93 GHz continuum peak, and the contours are 93 GHz continuum at \((0.025, 0.05, 0.1, 0.2, 0.4, 0.6, 0.8, 0.95)\times2.59~\mathrm{mJy~beam}^{-1}\). (b) Spectral energy distribution of the core plotted as the total flux density \(S_\nu\) (Table \ref{tab:cont_flux}) as a function of frequency \(\nu\). The curves are: synchrotron (dot-dashed), free-free (dotted), thermal dust (dashed), and total fit (full).\label{fig:cont3}}
\end{figure}

The spectral energy distribution (SED) of the galactic center (core) from 21 GHz to 350 GHz is plotted in Figure \ref{fig:cont3}(b). Also shown is a least-squares fit that includes contributions from three different types of emission: synchrotron (non-thermal), free-free, and dust (thermal). The fitting function is

\begin{equation}
S_\nu=K_\mathrm{s}\nu^{\alpha_\mathrm{s}}+K_\mathrm{ff}\nu^{\alpha_\mathrm{ff}}+K_\mathrm{d}\nu^{\alpha_\mathrm{d}},
\end{equation}
where \(K\) is the proportionality factor to be determined and \(\alpha\) is the corresponding spectral index. In order to simplify the fitting procedure, we fixed the spectral indices as \(\alpha_\mathrm{s}=-0.9\) (synchrotron), chosen to be consistent with 1.4 and 5 GHz measurements by \cite{Dah90} and typical values in star-forming galaxies \citep{Con92}, \(\alpha_\mathrm{ff}=-0.1\) (free-free), and \(\alpha_\mathrm{d}=3.8\) (dust; emissivity index \(\beta=1.8\)). Note that the SED exhibits a minimum at \(\sim100~\mathrm{GHz}\); this is the frequency domain where free-free emission is typically dominant because of its characteristically flat spectrum. The resulting fractions of the contributions at 93 GHz are \(S_\mathrm{s}:S_\mathrm{ff}:S_\mathrm{d}=0.22:0.72:0.06\). The fitting result is not unique, but for the adopted values of spectral indices, it indicates that the flux density of the core at 93 GHz can be dominated by free-free thermal emission from \ion{H}{2} regions followed by non-thermal emission from relativistic electrons accelerated in magnetic fields, and a smaller contribution from the thermal cooling of dust grains. This is in agreement with the previously established picture of a nuclear starburst that may host an embedded low-luminosity AGN. The SED in the core of NGC 1808 in this frequency range is similar to that in the central regions of the starburst galaxies M82 and NGC 253 (e.g., \citealt{Kle88,CK91,Ben15}).

The total flux of the 93 GHz continuum in the central region is \(S_\mathrm{93}(r<15\arcsec)=3.24\times10^{-2}\) Jy. For an electron temperature \(T_\mathrm{e}\), the production rate of Lyman continuum photons in \ion{H}{2} regions\footnote{Lyman continuum photons have energies higher than the ionization energy of hydrogen atoms (13.6 eV); they are emitted primarily by massive stars.}  can be calculated from (e.g., \citealt{Mat05})

\begin{equation}
\left(\frac{Q_\mathrm{Lyc}}{\mathrm{s}^{-1}}\right)=1.4\times10^{55}\left(\frac{S_\mathrm{93}}{\mathrm{Jy}}\right)\left(\frac{d}{10.8~\mathrm{Mpc}}\right)^2\left(\frac{\nu}{93~\mathrm{GHz}}\right)^{0.1}\left(\frac{T_\mathrm{e}}{10^4~\mathrm{K}}\right)^{-0.45}k_\mathrm{ff},
\end{equation}
where \(k_\mathrm{ff}\) is the ratio of free-free emission flux density to the total flux density at 93 GHz. Adopting \(T_\mathrm{e}=5000\) K, estimated for the central region of the starburst galaxy M82 (and comparable to the nucleus of the starburst galaxy NGC 253, where \(T_\mathrm{e}=3700\)-\(4500\) K \citep{Ben15}, and Galactic \ion{H}{2} regions where typically 5000-9000 K), and \(k_\mathrm{ff}=0.72\) from the fit above (assuming that the free-free fraction is the same across the whole region), we find \(Q_\mathrm{Lyc}=4.4\times10^{53}~\mathrm{s}^{-1}\). The result is comparable to that obtained for the central 500 pc region of M82 \citep{CK91} and the nucleus of NGC 253 \citep{Ben15}. Following the calculation by \cite{Mat05}, the measured \(Q_\mathrm{Lyc}\) is equivalent to \(\sim10^5\) stars with masses \(\gtrsim8~M_\sun\), which can become the progenitors of supernova explosions. The supernovae input mechanical energy into the ISM that drives the superwind discussed in sections \ref{Dc0} and \ref{Dc}.

\subsection{CO (3-2) intensity distribution}\label{Cb}

\begin{figure}
\epsscale{1.1}
\plotone{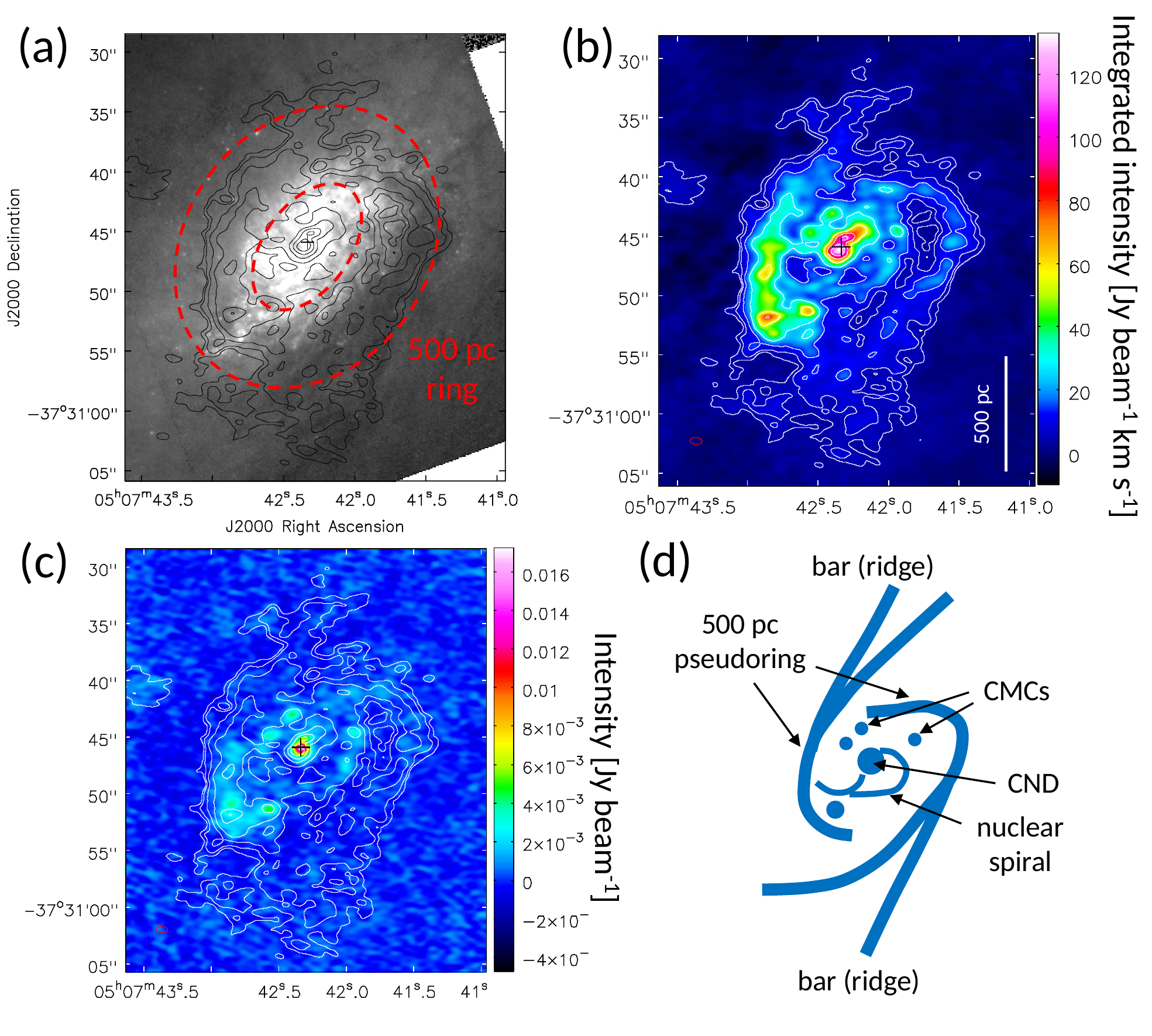}
\caption{(a) \(R\)-band HST image from the \emph{Hubble Legacy Archive} with CO (3-2) contours plotted at \((0.03, 0.05, 0.1, 0.2, 0.4, 0.6, 0.8)\times133.4~\mathrm{Jy~beam^{-1}~km~s^{-1}}\) (maximum). The plus sign marks the adopted galactic center (93 GHz continuum peak). The dashed red ellipses mark the approximate extent of the 500 pc ring. The major axes of the inner and outer ellipses are \(12\arcsec\) and \(25\arcsec\) long. (b) CO (3-2) integrated intensity. (c) CO (3-2) contours plotted over a 350 GHz continuum image. (d) Illustration of the main structures of molecular gas distribution: CND (circumnuclear disk), CMCs (central molecular clouds), and 500 pc pseudoring. The size of the synthesized beam is shown at the bottom left corner.\label{fig:co32}}
\end{figure}

Molecular gas distribution traced by the CO (3-2) line is shown in Figure \ref{fig:co32} as integrated intensity (moment 0 image) \(\mathcal{I}_\mathrm{CO}=M_0\equiv\Delta v\sum_i \mathcal{S}_i\), where \(\Delta v\) is the velocity resolution and \(\mathcal{S}_i\) is the flux density per beam solid angle (intensity in units Jy beam\(^{-1}\)) in a spectral channel \(i\). The total integrated flux in the central \(40\arcsec\) derived from the \(M_0\) image clipped at \(5~\sigma\) is \(I_\mathrm{CO(3-2)}=8855\) Jy km s\(^{-1}\). Also shown are optical (R band with an effective wavelength of 658 nm) and 350 GHz continuum images with CO (3-2) intensity contours. Optical observations have previously revealed a network of dust lanes across the central 1 kpc region, some of which appear to be extraplanar and extend outwards from the galactic center as far as 3 kpc (Figures \ref{fig:n1808} and \ref{fig:co32}(a)).

The CO (3-2) line is detected throughout the region, and especially in the central 100 pc (referred to as the circumnuclear disk; CND), where a maximum value of the integrated intensity of \(133.4~\mathrm{Jy~beam^{-1}~km~s^{-1}}\) is measured. The CND, somewhat similar to the Galactic Central Molecular Zone (CMZ), harbors the core shown in continuum images, that exhibits a nuclear starburst and a low-luminosity AGN. The CND is surrounded by a pseudoring (two molecular spiral arms) at a radius of \(\sim500~\mathrm{pc}\) from the center (hereafter the ``500 pc ring''; Figure \ref{fig:co32}). The overall structure is very similar to the distribution of CO (1-0) intensity at lower resolution (\(\sim2\arcsec\)) reported by \cite{Sal16}, who analyzed the radial distribution of molecular gas based on the CO (1-0) data. A closer look at the optical and CO (3-2) images in Figure \ref{fig:co32} suggests that the distribution of molecular gas is coincident with the dust lanes, especially in the 500 pc ring (panel a). Between the CND and the 500 pc ring, we confirm the presence of a nuclear (molecular) spiral pattern; owing to higher angular resolution, the data presented here reveal at least two spiral arms and a number of compact sources referred to as the central molecular clouds (CMCs). The CMCs reside in the same region (within \(R<500\) pc from the galactic center) as do the hot spots identified in the radio continuum (Figure \ref{fig:cont2}). In sections \ref{Da} and \ref{Db}, we derive the molecular gas mass in the CMCs and investigate their star formation activity. An illustration of the CND, CMCs, spiral arms, and 500 pc ring is shown in Figure \ref{fig:co32}(d).

Figure \ref{fig:co32}(c) shows a comparison between the CO (3-2) intensity (contours) and 350 GHz continuum (image) produced by thermal emission from dust grains. Although the dust continuum is detected to a lower spatial extent, there is similarity with the CO (3-2) intensity distribution; the emission is most prominent in the CND, where a peak of \(\mathcal{S}_\mathrm{350}^\mathrm{max}=17~\mathrm{mJy~beam}^{-1}\) is measured, as well as in CMCs, and partially in the 500 pc ring. The continuum emission is also weakly detected in the nuclear spiral arms.

\subsection{CO (3-2)/(1-0) line intensity ratio in the central 1 kpc}\label{Ce}

\begin{figure}
\epsscale{1}
\plotone{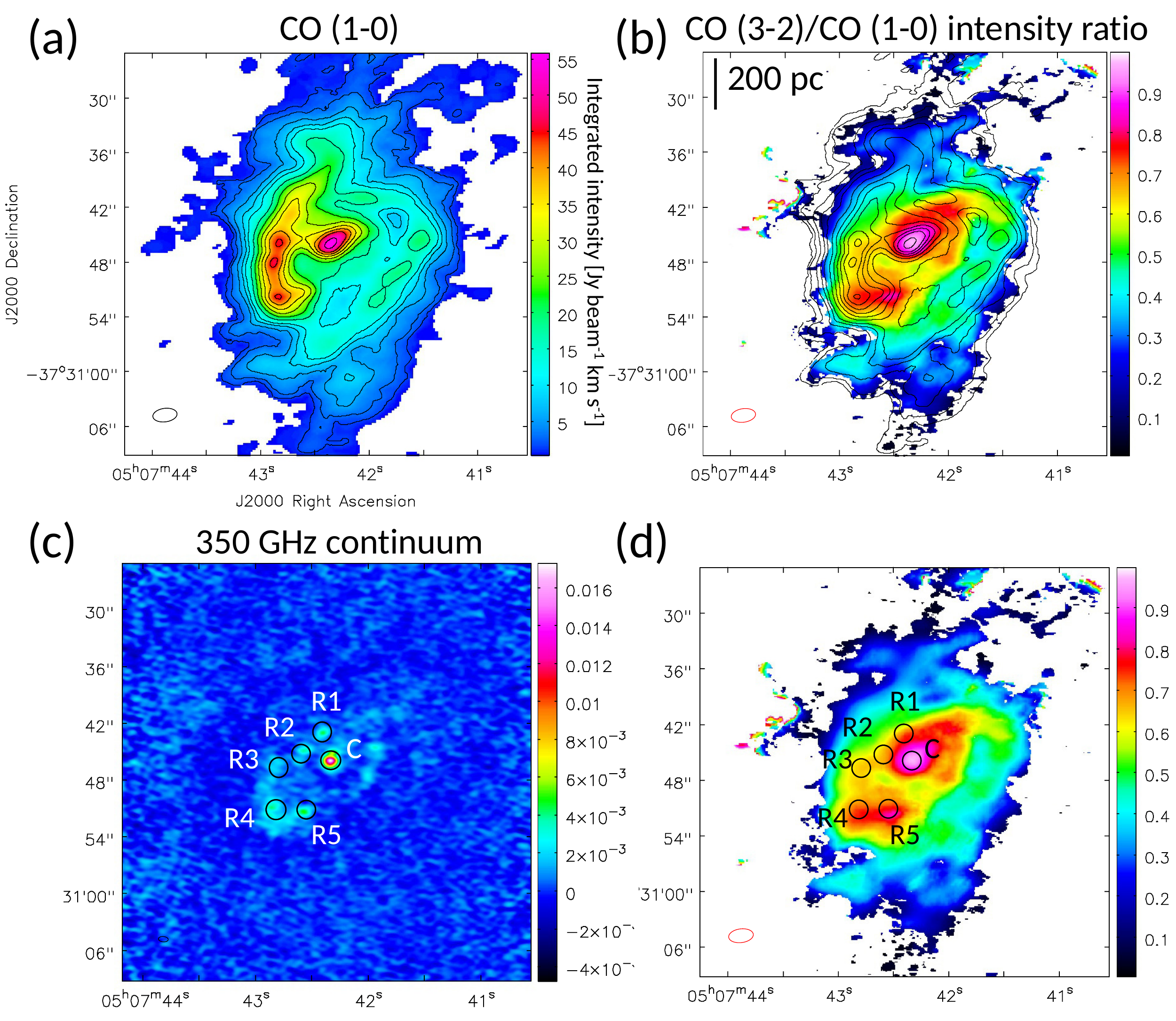}
\caption{(a) CO (1-0) integrated intensity of the combined 12m + ACA image. The contours are plotted at \((0.025, 0.05, 0.1, 0.15, 0.2, 0.3, 0.4, 0.5, 0.6, 0.7, 0.8, 0.95)\times55.9~\mathrm{Jy~beam^{-1}~km~s^{-1}}\) (maximum value). The image intensity is clipped at \(5~\sigma\) where \(1~\sigma=5\) mJy beam\(^{-1}\). (b) CO (3-2) to (1-0) integrated intensity ratio (each image in brightness units K km s\(^{-1}\)) derived after adjusting the CO (3-2) image to the angular resolution of CO (1-0), \(2\farcs67\times1\farcs48\) (\(\sim100\) pc). The contours are CO (1-0) as in panel (a). (c) 350 GHz image with investigated regions marked by circles (diameter 2\arcsec). (d) CO (3-2)/(1-0) intensity ratio image with investigated regions.\label{fig:linrat}}
\end{figure}

Figure \ref{fig:linrat} shows the integrated intensity of the CO (1-0) line corrected for missing flux (panel a), and the line intensity ratio of CO (3-2)/CO (1-0) derived in the central 1 kpc at a resolution of \(2\arcsec\) (\(\sim100\) pc) corresponding to that of the CO (1-0) image (panel b). The CO line ratio in Figure \ref{fig:linrat}(b) is defined as \(R_\mathrm{CO}\equiv W_\mathrm{CO(3-2)}/W_\mathrm{CO(1-0)}\), where \(W_\mathrm{CO}=\int T_\mathrm{b}dv\) is the integrated intensity (brightness temperature \(T_\mathrm{b}\) integrated over velocity \(v\)) in K km s\(^{-1}\). The image shows that the ratio is not well correlated with the integrated intensity of the CO (1-0) emission; it is highest toward the central \(1\arcsec\) (\(R_\mathrm{CO}=0.96\pm0.01\)), with a tendency to gradually decrease with galactocentric radius. However, \(R_\mathrm{CO}\) is relatively low (0.4-0.5) in the 500 pc ring, and only increases to values of 0.5-0.7 in regions of the 500 pc ring NW and SE of the galactic center. The ratio is lowest (\(<0.4\)) in the outermost parts of the ring, with the lowest measured values of \(\sim0.2\) beyond the ring. The ratio of integrated intensities in the central 1 kpc region is \(R_\mathrm{CO}(R<10\arcsec)\sim0.57\), consistent with previously reported single-dish measurements \citep{Aal94,Sal14}.

In panels (c) and (d), we mark the locations of six selected regions, including the center (C) and CMCs (labelled R1-R5), whose line ratio parameters are listed in Table \ref{tab6}. These are investigated because they are representative regions that exhibit detections in all tracers (93 GHz and 350 GHz continuum, and CO), useful for comparison and derivation of properties such as gas mass and star formation rate. The star formation properties of the regions are presented in section \ref{D}.

\begin{table}
\begin{center}
\caption{Molecular Gas Parameters in Selected Regions (Diameter \(2\arcsec\))}\label{tab6}
\begin{tabular}{lccccr}
\tableline\tableline
Region & Coordinates\tablenotemark{a} &  \(R_\mathrm{CO}\) & \(I_\mathrm{CO(1-0)}\) & \(\Delta V_\mathrm{CO(1-0)}\) \\
& \([\mathrm{arcsecond},-'']\) & & \([\mathrm{Jy~km~s^{-1}}]\) & \([\mathrm{km~s}^{-1}]\) \\
\hline
C & \(42.331,-45.877\) & \(0.935\pm0.025\) & 37.1 & \(151.1\pm2.0\)  \\
R1 & \(42.402,-42.909\) & \(0.686\pm0.063\) & 16.5 & \(89.0\pm3.9\) \\
R2 & \(42.590,-45.159\) & \(0.664\pm0.034\) & 23.0 & \(111.2\pm1.8\) \\
R3 & \(42.788,-46.658\) & \(0.641\pm0.009\) & 28.6 & \(97.4\pm1.7\) \\
R4 & \(42.815,-51.127\) & \(0.725\pm0.027\) & 30.0 & \(68.4\pm1.2\) \\
R5 & \(42.543,-51.068\) & \(0.763\pm0.042\) & 18.4 &  \(70.2\pm1.6\) \\
\tableline
\end{tabular}
\end{center}
\tablenotemark{a}{The expressed coordinates have common values of \((\alpha,\delta)=(05^\mathrm{h}07^\mathrm{m},-37\degr30\arcmin)\).}
\tablecomments{The \(R_\mathrm{CO}\) ratio was derived at the angular resolution of the CO (1-0) line (see panels b and d in Figure \ref{fig:linrat}).}
\end{table}

\subsection{Gas torus and core in the central 100 pc}\label{Cd}

\begin{figure}
\epsscale{1.1}
\plotone{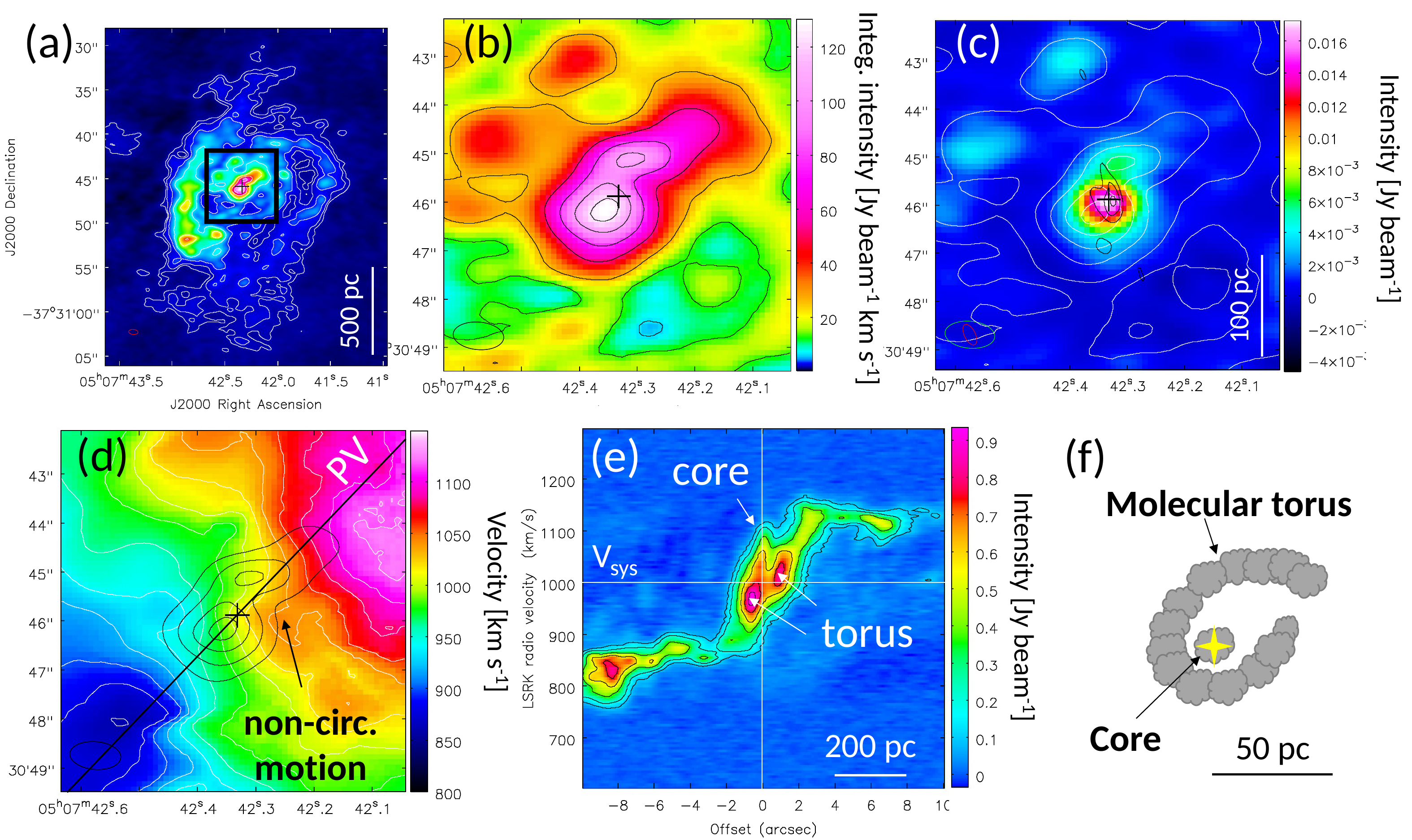}
\caption{(a) CO (3-2) integrated intensity (same as in Figure \ref{fig:co32}). The black plus sign marks the adopted galactic center (93 GHz continuum peak). (b) Enlargement of the black rectangle in panel (a). (c) 350 GHz continuum image, CO (3-2) intensity contours (white), and 21 GHz intensity contours (black) plotted as \((5, 10, 15)\times1~\sigma_\mathrm{K}\) (see Figure \ref{fig:cont1}). (d) Velocity field of molecular gas traced by CO (3-2) as moment 1 image. The white contours are plotted from 860 to 1120 km s\(^{-1}\) in steps of 20 km s\(^{-1}\) (LSR). The black contours are CO (3-2) intensity. An example of non-circular motions west of the core is marked by an arrow. (e) Position-velocity diagram plotted at position angle of \(316\arcdeg\) (black straight line in panel d); the slice width is \(1\arcsec\). The systemic velocity is \(V_\mathrm{sys}=998~\mathrm{km~s}^{-1}\). The contours are: \((0.1, 0.2, 0.4, 0.6, 0.8, 0.95)\times0.937~\mathrm{Jy~beam}^{-1}\). Note a double peak (torus) and an extended high-velocity component (core). (f) Illustration of the CND: the core is defined by the continuum peak; this is the location of the high-velocity core component in (e). \label{fig:torus}}
\end{figure}

The central 100 pc (CND) of the CO (3-2) intensity distribution is shown in Figure \ref{fig:torus} (panels b-d). In the CND, we show for the first time that CO (3-2) exhibits a double peak with a projected separation of \(1\farcs17\pm0\farcs07\). The separation was calculated by deriving the positions of the peaks from 2D Gaussian fitting, within circles of diameter \(0\farcs5\) centered at the brightest pixels. The uncertainty of the separation is set by the absolute positional accuracy in ALMA images with high signal-to-noise ratio (\(>10\)), which is estimated to be \(\sim5\%\) of the resolution, i.e., \(\sim0\farcs05\) for a beam size of \(\theta=1\arcsec\). For the adopted distance of \(d=10.8~\mathrm{Mpc}\), this yields a physical separation of \(61\pm4~\mathrm{pc}\). On the other hand, panel (c) shows that the continuum emission at 21 and 350 GHz reaches its maximum value between the CO (3-2) intensity peaks. The 21 and 350 GHz continuum peaks are separated by \(\sim0\farcs15\), which is comparable to the resolution of the low-frequency image. Thus, the continuum source is offset from both molecular gas concentrations (CO intensity peaks). Measured with respect to the mid-point between the CO peaks, the continuum source is offset by \(\sim14~\mathrm{pc}\) and closer to the higher-intensity (south-east) CO peak. The positions of the 350 GHz continuum peak and the south-east peak of CO (3-2), obtained from two-dimensional Gaussian fitting as described above, are \((\alpha,\delta)_\mathrm{345GHz}=(05^\mathrm{h}07^\mathrm{m}42\fs33567\pm0\fs00055,
-37\degr30\arcmin45\farcs93954\pm0\farcs00305)\) and \((\alpha,\delta)_\mathrm{CO(3-2)SE}=(05^\mathrm{h}07^\mathrm{m}42\fs35503\pm0\fs00055,
-37\degr30\arcmin46\farcs16905\pm0\farcs00504)\). Thus, the separation between the two peaks is estimated to be \(0\farcs32\pm0\farcs07\).

In order to investigate the motion of molecular gas in the CND, we created a velocity field image of the CO (3-2) line (first moment of intensity), defined as \(M_1=\Sigma_i \mathcal{S}_i v_i/M_0\), where \(M_0\) is the integrated intensity defined above. The result is shown in Figure \ref{fig:torus}(d). Note that although the bulk molecular gas exhibits rotational motion about the center, there is evidence of noncircular motions with deviations up to \(\sim50~\mathrm{km~s}^{-1}\). This was already discussed in \cite{Sal16} and \cite{Bus17} using CO (1-0) and infrared data of vibrationally excited H\(_2\), respectively, where it was found that noncircular motions are related to gas streaming motion possibly due to the nuclear bar. Inside the CND, our new CO (3-2) data reveal that the two CO-intensity peaks are at different line-of-sight velocities, suggesting circumnuclear rotation. In addition, the position-velocity diagram (PVD) plotted in Figure \ref{fig:torus}(e), at a position angle derived from dynamical fitting by \cite{Sal16}, confirms the presence of the double peak at velocities consistent with overall ``rigid-body'' rotation in the central 200 pc. Inspecting the new data, we also found that the velocity at the location of the core (continuum peak; offset \(0\arcsec\) in Figure \ref{fig:torus}(e)) deviates from this rigid-body rotation; there is a high-velocity component denoted by ``core'' in Figure \ref{fig:torus}(e).

In Figure \ref{fig:torus}(f), we illustrate a possible geometry of the molecular gas and continuum in the CND. The double peak of CO (3-2) intensity is modeled to be an inclined, rotating torus with a radius of \(r\sim30~\mathrm{pc}\). The torus is probably patchy and affected by noncircular motions and turbulence due to inflows from the nuclear spiral arms and nuclear bar, as well as outflows from the intense nuclear starburst/AGN (the nuclear outflow was discussed in \citealt{Sal16}). Recent numerical simulations show that a similar thick disk of cold neutral gas can naturally be formed by radiative feedback from a low-luminosity AGN and mechanical energy from supernova explosions in a nuclear starburst \citep{WSM16}. A similar structure has also been observed in the CMZ in the Galactic center (e.g., \citealt{Sof17}). Note that the high-velocity component in Figure \ref{fig:torus}(e) is spatially coincident with the continuum core in panel (c). We conclude that the ``core'' illustrated in panel (f) harbors both molecular gas and a luminous continuum source. This structure is very similar to that observed in the center of the Seyfert galaxy NGC 1068 \citep{GB16}.

The high-velocity component in the core observed in the PVD in Figure \ref{fig:torus}(e) can be produced by different phenomena: (1) high-velocity rotation around the central supermassive black hole or nuclear star cluster with a mass of \(\sim1\times10^7~M_\sun\) as proposed by \cite{Sal16}, possibly as an inclined or warped nuclear torus within the central \(\lesssim10\) pc, (2) nuclear streaming motion (e.g., \citealt{Bus17}), or (3) outflow, such as the one observed in the core of the Seyfert galaxy NGC 1068 \citep{GB14}. While the velocity is in agreement with rotation around a central massive object, the overall irregular motions west of the core (marked with an arrow in panel (d)) cannot exclude the possibility of nuclear streaming motions. Considering that the high-velocity component in the core is marginally resolved in the PVD in panel (e) and coincides with the continuum core, we prefer the scenario of a rotating nuclear torus within \(\lesssim10~\mathrm{pc}\). This picture is supported by mid-infrared observations which also suggest that the core may harbor an edge-on torus (column density \(N_\mathrm{H}\sim10^{24}\) cm\(^{-3}\)) within the central 1 pc, blocking high-energy radiation from an active nucleus \citep{Sale13}. This compact MIR source may correspond to the core revealed in our CO (3-2) data. Note that this structure in the core is much smaller (\(\lesssim10\) pc) than the molecular gas torus (radius 30 pc). A more detailed picture of the geometry and kinematics of the central 10 pc, requires new high-resolution observations that are now becoming possible with ALMA. Observations at \(\lesssim1\) pc resolution may allow us to measure the ratio of velocity dispersion to rotational velocity in the core and study its dynamics.

\section{Star formation}\label{D}

\subsection{ISM mass from 350 GHz dust continuum}\label{Da}

Recently, optically thin dust continuum at 350 GHz (Rayleigh-Jeans regime) has been used as a robust estimator of the total ISM (HI + H\(_2\)) mass (e.g., \citealt{Sco14,Sco15}). The observed flux density is directly proportional to the dust temperature \(T_\mathrm{d}\) and dust mass \(M_\mathrm{d}\) as \(S\propto\kappa_\mathrm{d}T_\mathrm{d}\nu^2 M_\mathrm{d}/d^2\), where \(\kappa_\mathrm{d}\) is the dust opacity per unit mass of dust, and \(d\) is the distance to the galaxy. \cite{Sco14} estimated that a large fraction of dust by mass has a temperature of \(T_\mathrm{d}=20\mathrm{-}40\) K in a sample of nearby star-forming galaxies. Furthermore, the authors empirically calibrated a mass-to-light conversion factor for dust so that the opacity and dust mass are not required to derive the total ISM mass.

Using equation (3) from \cite{Sco15}, we calculated the ISM mass of the compact sources C and R1-R6 from Figure \ref{fig:linrat} and Table \ref{tab6} in the central region. Since \(T_\mathrm{d}\) can be higher in the nuclear regions compared to large-scale average values, we calculated the ISM mass for dust temperatures of 20 and 100 K. For instance, \(T_\mathrm{d}\) was estimated to be of the order of 100 K in the double nucleus of Arp 220 \citep{Mat09,Wil14}. The results are given in Table \ref{tab6c}: all investigated regions have masses in the range \(10^6\lesssim M_\mathrm{ISM}/M_\sun\lesssim10^7\). Also listed are molecular gas masses (including He and other elements) derived from the CO (1-0) integrated flux (Table \ref{tab6}) by using a relatively low conversion factor \(X_\mathrm{CO}=0.8\times10^{20}~\mathrm{cm^{-2}(K~km~s^{-1})^{-1}}\) from \cite{Sal14} and the equation (e.g., \citealt{Bol13b}):

\begin{equation}\label{molmass}
\left(\frac{M_\mathrm{mol}}{M_\sun}\right)=1.05\times10^4\left(\frac{X_\mathrm{CO}}{2\times10^{20}~\mathrm{cm^{-2}(K~km~s^{-1})^{-1}}}\right)\left(\frac{d}{\mathrm{Mpc}}\right)^2\left(\frac{I_\mathrm{CO}}{\mathrm{Jy~km~s}^{-1}}\right).
\end{equation}
The two measurements agree within the uncertainty of \(T_\mathrm{d}\) and \(X_\mathrm{CO}\) (a factor of few), revealing that the investigated regions harbor massive molecular gas reservoirs.

\begin{table}
\begin{center}
\caption{Flux Densities, ISM Masses, Star Formation Rates and Efficiencies in Selected Regions (Diameter \(2\arcsec\))}\label{tab6c}
\begin{tabular}{lccccccc}
\tableline\tableline
Region & \(S_\mathrm{350}\) & \(S_\mathrm{93}\) & \(M_\mathrm{ISM}\) (\(T_\mathrm{d}=20\) K) & \(M_\mathrm{ISM}\) (\(T_\mathrm{d}=100\) K) & \(M_\mathrm{mol}\) (\(X_\mathrm{CO}\)) & SFR\(_\mathrm{ff}\) & \(\mathrm{SFE_{ff}}\equiv\mathrm{SFR_{ff}}/M_\mathrm{mol}\) \\
& \([\mathrm{mJy}]\) & \([\mathrm{mJy}]\) & \([\times10^7~M_\sun]\) & \([\times10^6~M_\sun]\) & \([\times10^7~M_\sun]\) & \([M_\sun~\mathrm{yr}^{-1}]\) & \([\times10^{-8}~\mathrm{yr}^{-1}]\) \\
\hline
C & 17.5 & 2.01 & \(3.47\) & \(4.82\) & \(1.82\) & 0.28 & 1.5 \\
R1 & 5.25 & 0.692 & \(1.04\) & \(1.45\) & \(0.808\) & 0.096 & 1.2 \\
R2 & 5.51 & 1.00 & \(1.09\) & \(1.52\) & \(1.13\) & 0.14 & 1.2 \\
R3 & 4.85 & 0.513 & \(0.962\) & \(1.34\) & \(1.40\) & 0.071 & 0.51 \\
R4 & 8.80 & 0.917 & \(1.74\) & \(2.42\) & \(1.47\) & 0.13 & 0.88 \\
R5 & 5.97 & 0.845 & \(1.18\) & \(1.64\) & \(0.901\) & 0.12 & 1.3 \\
\tableline
\end{tabular}
\end{center}
\tablecomments{The fluxes were derived after adjusting the angular resolution of the continuum images to that of CO (1-0).}
\end{table}

\subsection{Star formation rate and efficiency from 93 GHz continuum}\label{Db}

The star formation rate (SFR)\footnote{Defined as the mass of interstellar gas that forms stars per unit time, \(\Delta M_*/\Delta t\).} in the CND and regions R1-R5 can be estimated from the band 3 continuum assuming that the SED is dominated by free-free emission at 93 GHz. Following \cite{Mur11}, the SFR can be expressed as

\begin{equation}
\left(\frac{\mathrm{SFR}}{M_\sun~\mathrm{yr}^{-1}}\right)_\mathrm{ff}=4.6\times10^{-28}\left(\frac{T_\mathrm{e}}{10^4~\mathrm{K}}\right)^{-0.45}\left(\frac{\nu}{\mathrm{GHz}}\right)^{0.1}\left(\frac{L_\mathrm{ff}}{\mathrm{erg~Hz^{-1}~s^{-1}}}\right),
\end{equation}
where \(T_\mathrm{e}\) is the electron temperature in the ionized gas, and \(L_\mathrm{ff}\) is the luminosity derived as \(L_\mathrm{ff}=4\pi d^2 S_\mathrm{93}\), assuming spherically symmetric ionized gas regions. We adopt \(T_\mathrm{e}=5000\) K (see section \ref{Cab}), insert \(\nu=93\) GHz, and calculate the SFR using the luminosities derived from flux densities \(S_\mathrm{93}\) listed in Table \ref{tab6c}.

The total SFR in the CND and five selected regions is \(\mathrm{SFR}\sim0.8~M_\sun~\mathrm{yr}^{-1}\), approximately one fifth of the total \(\mathrm{SFR}(r<15\arcsec)\sim4.5~M_\sun~\mathrm{yr}^{-1}\) in the central starburst region derived from the total flux of the 93 GHz emission, and consistent with previous work \citep{KSG94}. By comparison, the SFR in the entire disk of the Milky Way galaxy averaged over the past 3 million years is estimated to be \(\sim1.3~M_\sun~\mathrm{yr}^{-1}\) \citep{MR10}; the central region of NGC 1808 is undergoing a starburst episode where the SFR in the CND and regions R1-R5 alone is comparable to the total SFR in the Galaxy. The SFR in the central 1 kpc is also similar to that derived for M82 applying the same method (\(7~M_\sun~\mathrm{yr}^{-1}\) in the case that 3 mm continuum is due to free-free emission; \citealt{Sala14}). The SFR in the CND (nuclear starburst) is between 0.20 and 0.28 \(M_\sun~\mathrm{yr}^{-1}\), where the lower limit is set by the estimated contribution of free-free emission in the galactic center (section \ref{Cab}). A higher electron temperature of \(T_\mathrm{e}=10^4\) K would result in \(\approx0.7\) times smaller SFR. The result is consistent with \(\sim0.2~M_\sun~\mathrm{yr}^{-1}\) derived within a smaller region of \(0\farcs75\) by \cite{Bus17} from the Br\(\gamma\) line of hydrogen, and by \cite{Esq14}, who derived the SFR within \(0\farcs35\) from the polycyclic aromatic hydrocarbon (PAH) emission at 11.3 \micron. It is also comparable to the SFR of \(0.53~M_\sun~\mathrm{yr}^{-1}\) in the nucleus derived by \cite{Kot96} from low-frequency radio and near-infrared data.

The star formation efficiency (SFE), defined as \(\mathrm{SFE}\equiv\mathrm{SFR}/M_\mathrm{mol}\), derived by using a constant conversion factor \(X_\mathrm{CO}\) (Table \ref{tab6c}), is highest in the CND and a factor of two-three lower in the R1-R5 regions. The upper limit of SFE in the CND is set by \(M_\mathrm{ISM}(T_\mathrm{d}=100~\mathrm{K})\) and therefore unlikely to be higher than about twice the derived value. On the other hand, the lower limit set by \(M_\mathrm{ISM}(T_\mathrm{d}=20~\mathrm{K})\) suggests that SFE cannot be lower than about one third of the derived value. Therefore, the SFE is of the same order of magnitude in the CND and R1-R5 regions unless there are extreme variations in the free-free emission contribution among the regions.

Two of the regions investigated here (R1 and R2) roughly correspond to the regions denoted by R2 and R3 in \cite{Bus17}, respectively, though the aperture size is different. The SFE values derived here for the CND and regions R1-R5 out to the 500 pc ring reside close to the Kennicutt-Schmidt law, expressed as \(\Sigma_\mathrm{SFR}\propto \Sigma_\mathrm{mol}^n\), where \(\Sigma\) is the surface density, and \(n\) is an index of the order \(\sim1\) (e.g., \citealt{Ken98}). All the regions in Table \ref{tab6c} exhibit depletion times\footnote{Depletion time is defined as the inverse of SFE, i.e., the time needed for all gas to be converted to stars for a given SFR: \(t_\mathrm{dep}\equiv 1/\mathrm{SFE}\).} in the range of \(t_\mathrm{dep}=10^7\)-\(10^8\) yr, indicating high SFE compared to the average values in the disks of spiral galaxies (\(t_\mathrm{dep}\sim2\times10^9\) yr; \citealt{Big08,Ler08}).

The total mass of the molecular gas in the central starburst region is \(M_\mathrm{mol}=9.7\times10^8~M_\sun\), derived from CO (1-0) data (corrected for short baselines) using Equation \ref{molmass}. For the same region, it was shown above that \(\mathrm{SFR}\sim4.5~M_\sun~\mathrm{yr}^{-1}\). The resulting depletion time averaged over the central 1 kpc starburst region is \(t_\mathrm{dep}(R<1~\mathrm{kpc})\sim2\times10^{8}\) yr, again an order of magnitude shorter than the average value in the disks of spiral galaxies (i.e., consumption rate of 100\% during \(10^8\) yr in the language of discussion in \cite{Big08}), although consistent with starburst galaxies \citep{Ken98}.

\section{Molecular gas outflows}\label{F}

\subsection{Kinematics of the nuclear outflow}\label{Dc0}

\begin{figure}
\epsscale{1.1}
\plotone{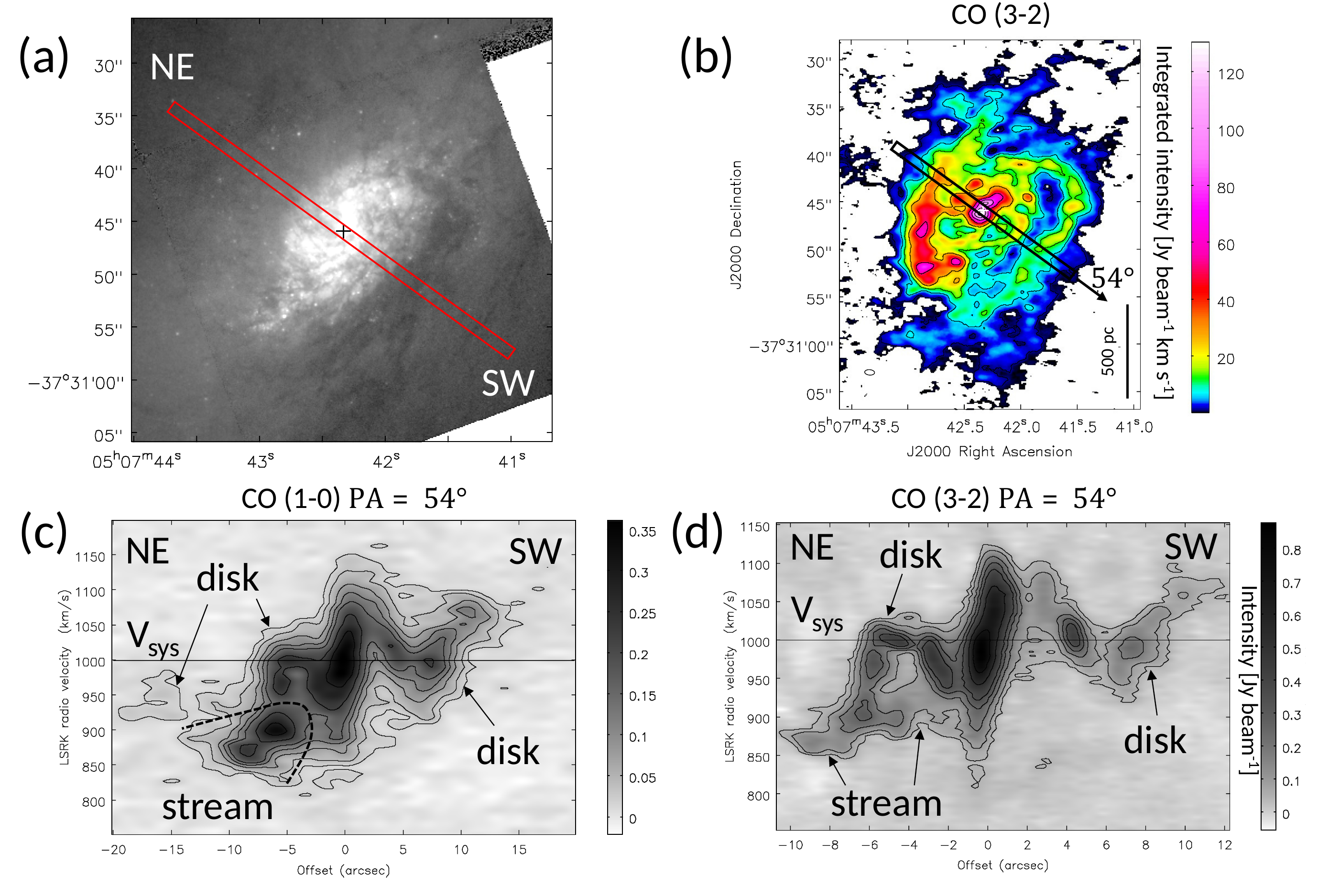}
\caption{(a) Position-velocity diagram (PVD) slit plotted on an \(R\)-band image along the minor galactic axis (\(\mathrm{PA}=54\arcdeg\)). (b) CO (3-2) image with a PVD slit. (c) CO (1-0) PVD. (d) CO (3-2) PVD in the inner region. Disk and stream (outflow) components are indicated. The contours are plotted at \((0.05, 0.1, 0.2, 0.3, 0.4, 0.6, 0.8)\times0.369\) Jy beam\(^{-1}\) for CO (1-0) and \((0.05, 0.1, 0.2, 0.3, 0.4, 0.6, 0.8)\times0.878\) Jy beam\(^{-1}\) for CO (3-2); the widths of the PVD slits are 5 pixels for CO (1-0) and 9 pixels for CO (3-2) (\(\simeq1\arcsec\)).\label{fig:pvdkin}}
\end{figure}

To study the distribution and kinematics of the nuclear outflow gas reported in \cite{Sal16}, it is important to resolve the outflow component of the CO emission from the rest of the central 1 kpc disk. Since the outflow and the disk overlap along the line of sight, we make maps in position-velocity space in the direction of the minor galactic axis and a conspicuous dark dust lane, where the outflow is expected to be recognized as CO line splitting offset from the velocity of galactic rotation (Figure \ref{fig:pvdkin} panels a and b). The position angle of the minor axis, \(\mathrm{PA}=54\arcdeg\), was derived from dynamical fitting of the rotation of the 500 pc ring and brightness distribution of the stellar bulge, where a position angle of the major axis was found to be 324\arcdeg \citep{Sal16}.

A molecular gas stream that largely deviates from circular motion is present in CO (1-0) and (3-2) emission in the position-velocity diagram (PVD) of Figure \ref{fig:pvdkin} (panels c and d) at LSR velocities of \(\sim900\) km s\(^{-1}\) in the NE and \(\sim1100\) km s\(^{-1}\) in the SW; the SW part is much fainter but visible in CO (1-0) data as low-intensity emission. The orientation of the NE side is toward the observer, thus expected to be blueshifted for an extraplanar gas outflow. The offset velocity component is not likely an inflow in the galactic plane (coplanar motion) because the line splitting would indicate that inflow orbits with radial velocities of the order of \(\sim100\) km s\(^{-1}\) cross with circular orbits all the way to the CND region, whereas hydrodynamical simulations of barred galaxies yield radial velocity amplitudes of the order of \(\lesssim50\) km s\(^{-1}\) with respect to the systemic velocity \(V_\mathrm{sys}\) (e.g., \citealt{Ren15}). For instance, note that the line width at offset \(r=-5\arcsec\) at \(\mathrm{PA}=54\arcdeg\) is nearly 150 km s\(^{-1}\). Assuming that the stream is perpendicular to the galactic plane (i.e., gas outflow), its velocity is \(|V_\mathrm{LSR}-V_\mathrm{sys}|/\cos{i}\approx180\) km s\(^{-1}\), where \(V_\mathrm{sys}=998\) km s\(^{-1}\) and \(i=57\arcdeg\) is the inclination of the galactic disk. Note that the CO (3-2) PVD shows that most of the CO gas in the region is not part of the stream, but resides in the galactic plane (\(V_\mathrm{LSR}\approx V_\mathrm{sys}\pm50\) km s\(^{-1}\)). The origin of the outflow is most likely the nuclear starburst, possibly coexisting with a weak AGN. The power source and the observed outflow velocities are consistent with recent theoretical studies, e.g., by \cite{Roy16}.

Next, we investigate the kinematics of the stream, using the new CO (1-0) and (3-2) data. We derive the velocity gradient by measuring the line-of-sight velocity of the stream component and correcting for the galactic inclination assuming that the bulk of the material is ejected perpendicular to the galactic plane, along the galactic minor axis (observed in a PVD through the galactic center; Figure \ref{fig:pvdkin}(d)). The adopted direction coincides with the orientation of polar dust lanes that appear to emerge from the CND region (Figure \ref{fig:co32}(a)), and extended [\ion{N}{2}] and H\(\alpha\) emission \citep{SB10}.

Figure \ref{fig:pvdkin}(d) shows that the stream component is clearly separated (velocity splitting \(\sim100\) km s\(^{-1}\)) from the circularly rotating disk component between offsets \(r=-2\arcsec\) (CND) and \(r=-8\arcsec\), and that the line-of-sight velocity \(|V_\mathrm{LSR}-V_\mathrm{sys}|\) of the stream is increasing with distance from the galactic center. If the molecular gas is flowing perpendicular to the galactic plane in the dark lanes and the outflow is replenished with gas at a constant ejection speed, the measured velocity gradient implies acceleration with distance from the galactic center. The velocity changes from \(r=-3\arcsec\) to \(r=-9\arcsec\), i.e., over \(6\arcsec\) equivalent to \(\sim600\) pc, from \(V_\mathrm{LSR}(r=-3\arcsec)\simeq930\) km s\(^{-1}\) to \(V_\mathrm{LSR}(r=-9\arcsec)\simeq865\) km s\(^{-1}\). Beyond offset \(r=-10\arcsec\) (deprojected height \(\sim1\) kpc), the signal-to-noise ratio of the CO (3-2) data cube is too low to make significant estimates. The CO (1-0) PVD in Figure \ref{fig:pvdkin}(c) traces the outflow up to \(V_\mathrm{LSR}(r=-15\arcsec)\simeq900\) km s\(^{-1}\), showing no sign of higher velocities.

Corrected for galactic inclination and systemic velocity, the outflow velocity derived from the CO (3-2) PVD appears to rise from \(\sim125\) km s\(^{-1}\) at \(h\sim290\) pc to \(\sim244\) km s\(^{-1}\) at \(h\sim600\) pc, yielding a velocity gradient of \(\Delta v/\Delta h\sim0.4\) km s\(^{-1}\) pc\(^{-1}\). As discussed in \cite{Sal16}, if the molecular gas does not gain additional momentum beyond this height, the bulk of the outflow gas is likely bound to the galaxy, unable to climb the gravitational well (minimum escape velocity \(\sim300\) km s\(^{-1}\) at galactocentric distance 1 kpc, derived from the total mass within that radius and assuming spherical symmetry), and expected to fall down onto the galactic disk.

Interestingly, a polar velocity gradient in NGC 1808 was also reported by \cite{Phi93}, who detected a neutral gas outflow up to \(\sim3\) kpc above the galactic plane from absorption and emission features of the \ion{Na}{1} D line. The maximum outflow velocities were estimated to be of the order of 400 to 700 km s\(^{-1}\), derived along a dark lane at \(\mathrm{PA}=50\arcdeg\) at a projected galactocentric distance of \(30\arcsec\) adopting the galactic inclination \(i=57\arcdeg\).  Although this is larger than our derived values of \(\sim244\) km s\(^{-1}\) at the projected distance of \(10\arcsec\), the data are consistent with each other because of a similar linear increase of velocity. It is therefore possible that the diffuse neutral gas detected in the \ion{Na}{1} line is accelerated to higher velocities deeper into the halo; it was found that the velocity no longer rises beyond a projected radius of \(\sim50\arcsec\). On the other hand, the nuclear outflow velocity of ionized gas traced by broad wings of [\ion{N}{2}] and [\ion{S}{2}] emission lines was estimated to be of the order of \(\gtrsim700\) km s\(^{-1}\). CO emission was not detected at these high velocities; if there is such fast neutral gas outflow, its mass must be negligible compared to the bulk molecular gas in the starburst region.

The apparent acceleration is also comparable to that recently found in the starburst galaxy NGC 253 (indication of a velocity gradient of 1 km s\(^{-1}\) pc\(^{-1}\) over an outflow size of 300 pc; \citealt{Wal17}), and possibly reflects the general properties of molecular gas in starburst-driven winds. Recent theoretical works suggest that cool gas can be accelerated initially by ram pressure of hot expanding gas, disrupted by instabilities, and recombine into molecules as it decelerates farther into the halo (e.g., \citealt{Tho16}). The fact that no molecular gas was detected in the nuclear outflow at high velocities of \(\sim700\) km s\(^{-1}\) reported by \cite{Phi93} could indicate that the fast nuclear outflow is devoid of molecules, i.e., that molecular clouds are ejected at lower velocities. Reformation of molecules at higher galactocentric distances may also be possible, since starburst-driven winds can contain dust, necessary for efficient (re)production of H\(_2\), as evident from the extraplanar dark lanes in NGC 1808 extending at least \(\sim3\) kpc above the galactic plane (Figure \ref{fig:n1808}), and, e.g., from evidence of a halo of cool dust in M82 (e.g., \citealt{Rou10}). On the other hand, cold clouds entrained in flows of hot gas may survive for a while as dense clumps or ``cloudlets'' in filamentary structure where Kelvin-Helmholz instabilities are less destructive (e.g., \citealt{Coo09,SB17}).

\begin{figure}
\epsscale{1}
\plotone{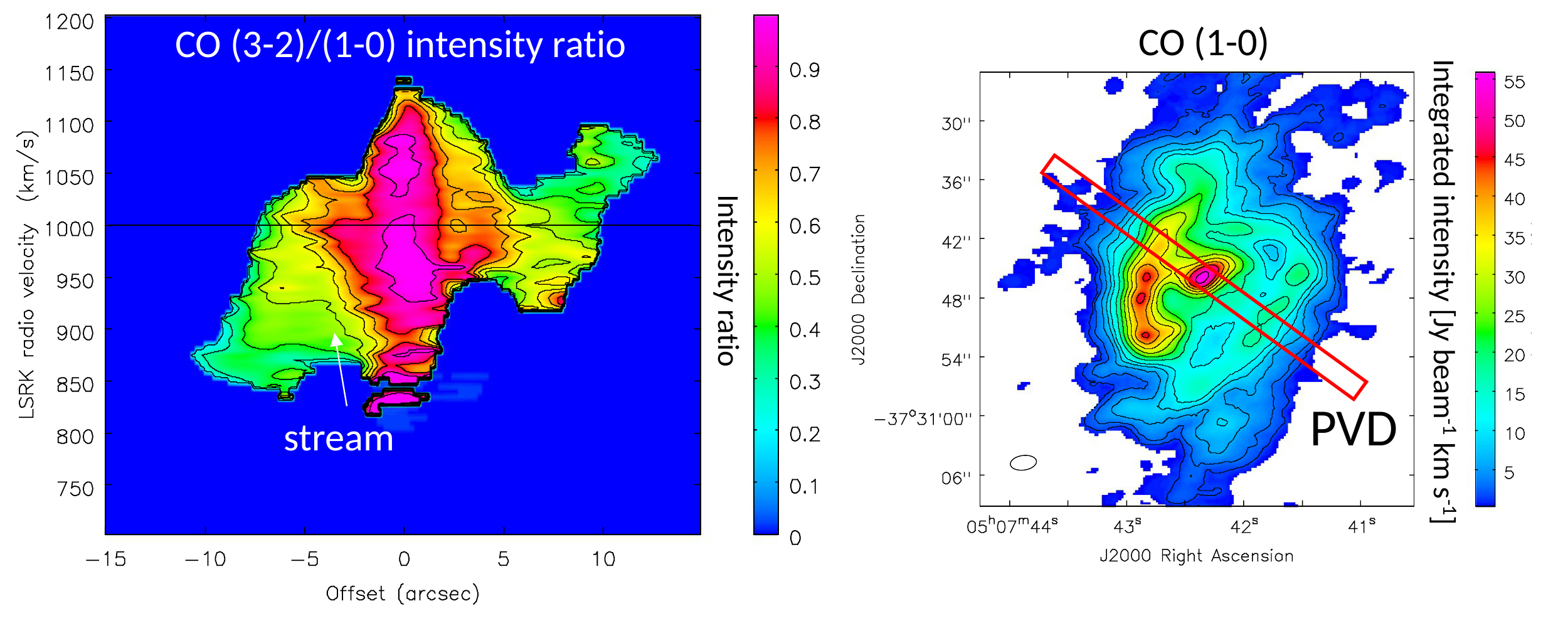}
\caption{Left) CO (3-2) to (1-0) integrated intensity ratio; contours are plotted at \(0.2, 0.3, 0.4, 0.5, 0.6, 0.7, 0.8, 0.95\). A nuclear outflow component is marked with an arrow. Right) CO (1-0) image with a PVD slice along the minor axis.\label{fig:pvdout}}
\end{figure}

Lastly, we present a CO (3-2)/CO (1-0) line intensity ratio image (\(R_\mathrm{CO}\)) of the nuclear outflow in Figure \ref{fig:pvdout}, adjusted to the angular resolution of the CO (1-0) image. The ratio image \(R_\mathrm{CO}\) in position-velocity space (shown in the left panel) exhibits variation in the range \(0.2<R_\mathrm{CO}<0.5\) in the outflow region. This is lower by a factor of \(\sim2\) compared to the CND and CMC regions, suggesting different physical conditions. Note also that the \(R_\mathrm{CO}\) ratio is decreasing from 0.5 to 0.2 along the velocity gradient shown in Figure \ref{fig:pvdkin}, indicating a gradual change of physical conditions in the molecular outflow as it recedes from the galactic center. A similar trend of decreasing \(R_\mathrm{CO}\) in extraplanar gas has been observed with single-dish telescopes in the outflow of the starburst galaxy M82 \citep{SC01,WWS05,Sal13}. In particular, the outflow in NGC 1808 displays a ratio gradient from \(R_\mathrm{CO}\sim1\) in the galactic center to \(R_\mathrm{CO}\sim0.3\) at a height of 1 kpc, which is nearly the same as the one measured in the outflow direction of M82 \citep{Sal13}. As in M82, this may imply low (beam-averaged) gas density in the outflow material, but detailed radiative transfer calculations that involve more lines would be necessary to probe the variation of physical conditions in the outflow.

Further observations will help us understand whether accelerated cold outflows are common in starburst-driven winds, what their maximum velocities are, and how the cold ISM phase in the winds evolves with galactocentric distance.

\subsection{Molecular gas outflows from the 500 pc ring}\label{Dc}

The new CO (1-0) data, corrected for the missing flux, also show evidence of molecular outflows in other regions in the central 1 kpc. Most of the new features were detected after the visibilities sampled by the ACA in the central region of the \((u,v)\) plane (short baselines) were added to the data set.

\begin{figure}
\epsscale{1.1}
\plotone{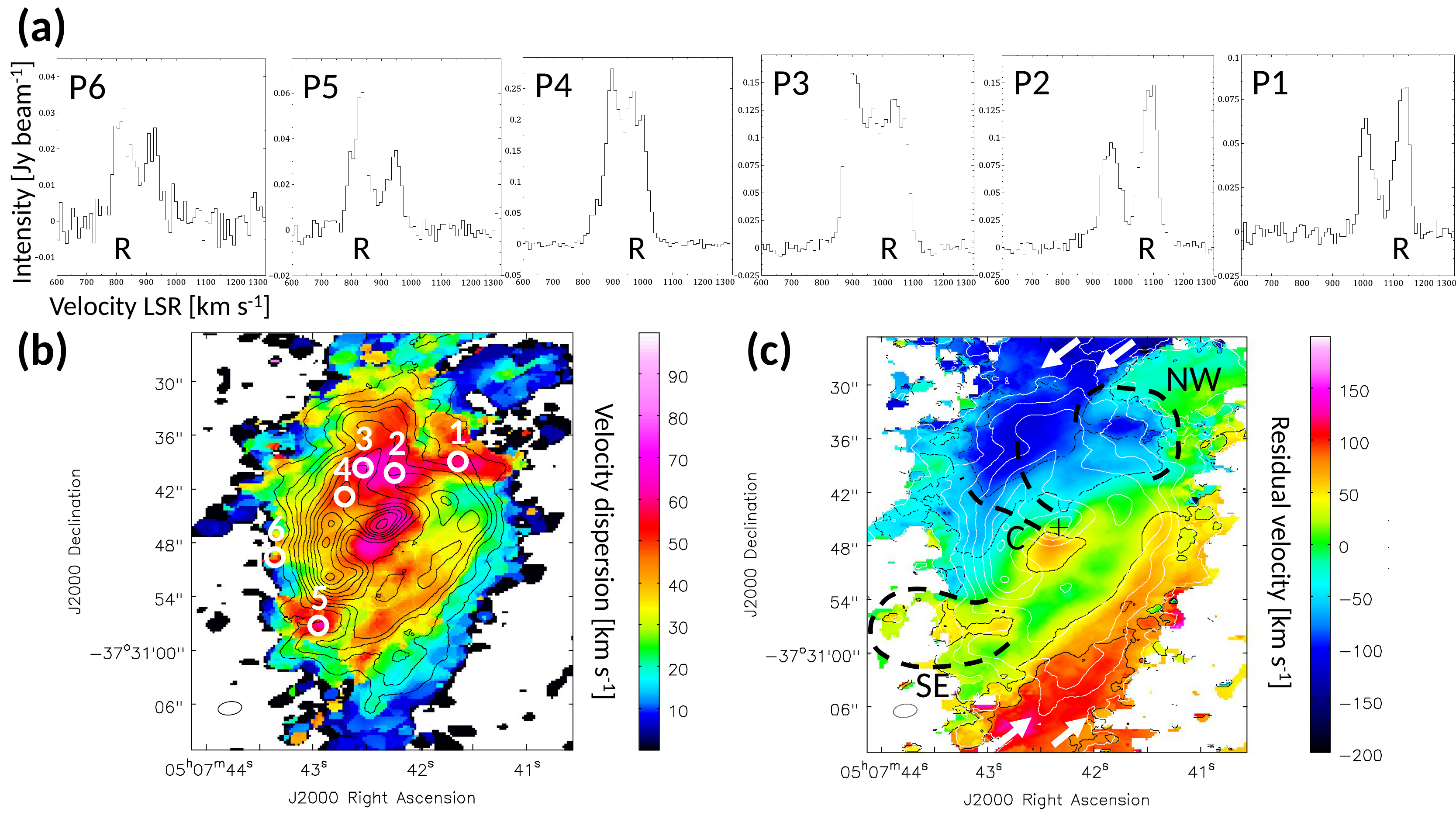}
\caption{(a) Spectra toward regions with large velocity dispersion shown on a moment 2 image of CO (1-0) in panel b. The positions 1-6 exhibit double and triple peaks, where one component (denoted by ``R'' in the spectra) belongs to the gas in circular rotation around the galactic center. The component separations are 100-150 km s\(^{-1}\) (b) CO (1-0) velocity dispersion with integrated intensity contours. (c) Residual velocity (model velocity field subtracted from CO (1-0) moment 1 image). Black contours are plotted at \(-80, -40, 40, 80\) km s\(^{-1}\); white contours are CO (1-0) intensity. Outflow regions are illustrated with black dashed lines (C, NW, SE). Non-circular motions in the ridges of the large-scale bar are indicated by white arrows.\label{fig:dv}}
\end{figure}

In order to search for extended outflows, we derived a velocity dispersion map (moment 2 of CO (1-0) intensity), defined as \(\sigma_\mathrm{disp}\equiv\sqrt{\Sigma\mathcal{S}_i(v_i-M_1)^2/M_0}\), where \(M_0\) and \(M_1\) are moment 0 and moment 1 images, respectively. The resulting \(\sigma_\mathrm{disp}\) image is shown in Figure \ref{fig:dv}(b). The image reveals that \(\sigma_\mathrm{disp}\) is large (typically 30-50 km s\(^{-1}\)) in the central 1 kpc starburst region, with local enhancements, especially in the CND. The CO (1-0) spectra of some regions in the 500 pc ring with \(\sigma_\mathrm{disp}\gtrsim60\) km s\(^{-1}\) (denoted by 1-6 in Figure \ref{fig:dv}(a)) show that the enhanced velocity dispersion is accompanied by at least two different emission components. The components are clearly separated, with total line widths of up to 200 km s\(^{-1}\) in some regions (see also Figure \ref{fig:pvdkin}(d)). These line widths are too large to be explained by velocity dispersion within gas clouds, but only as separate gas flows along the line of sight. Since one of the components can be explained by planar circular motion around the galactic center (denoted by ``R'' in the spectra), the offset components can be tracing gas outflows from the 500 pc ring.

The gas flows can be analyzed using the moment 1 image of the CO (1-0) data cube corrected for missing flux. We derived a model velocity field of pure circular motion in the galactic disk in the central 1 kpc using the rotation curve from Figure 20 of \cite{Sal16}, and subtracted it from the CO (1-0) moment 1 image. In the calculation of the model image, we adopted a systemic velocity of \(V_\mathrm{sys}=985\) km s\(^{-1}\), appropriate for the global kinematics of NGC 1808, position angle \(\mathrm{PA}=324\arcdeg\), and inclination \(i=57\arcdeg\). The result is shown in Figure \ref{fig:dv}(c).

The residual velocity image confirms striking non-circular motions (velocity residuals of the order 50-100 km s\(^{-1}\)) in the regions beyond the 500 pc ring N and S of the center, where the ridges of the large-scale bar connect to the 500 pc ring (white arrows in panel c); the velocity field here shows evidence of global inflow motions driven by bar dynamics. However, the locations of some of the regions with high velocity dispersion, such as 1 and 5 in panel (b), equivalent to NW and SE in panel (c), are not along the bar ridges; their origin is more easily explained as extraplanar motion. All outflow regions are marked by sharp velocity turns and line splittings with respect to the expected circular rotation that arise due to the presence of multiple velocity components (panel a in Figure \ref{fig:dv}).

The velocity field shows that the kinematics of the 500 pc ring is dominated by circular rotation (residuals \(\lesssim50\) km s\(^{-1}\)). Therefore, it is likely that the gas clouds from the bar ridges gradually accrete onto the 500 pc ring. This result will be used in section \ref{E} in the discussion of the evolution of molecular clouds.

\begin{figure}
\epsscale{1.1}
\plotone{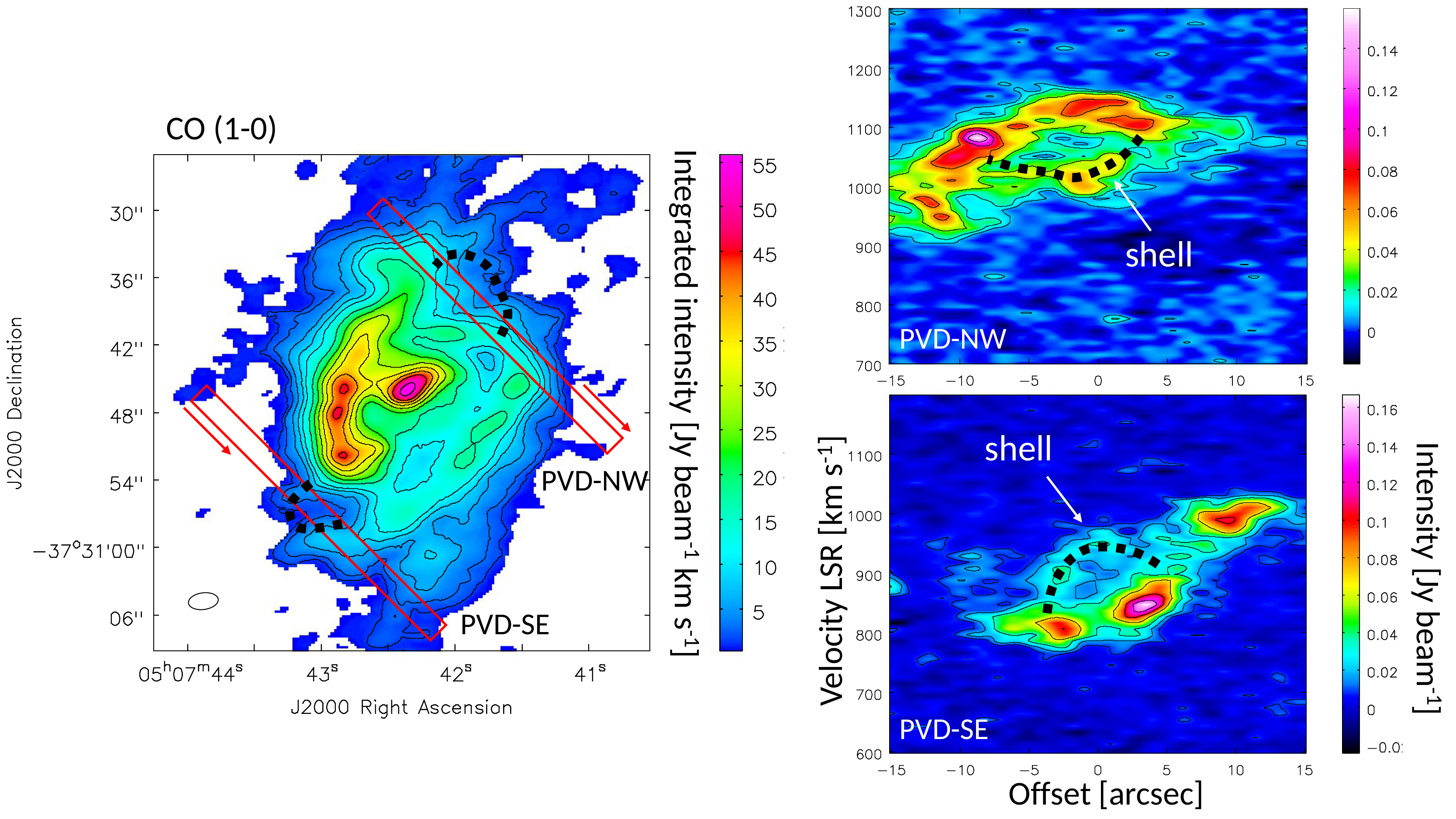}
\caption{Outflows in the 500 pc ring revealed by CO (1-0) line splitting in PVDs with a maximum velocity width of \(\Delta v_\mathrm{s}\sim150\) km s\(^{-1}\). Black dotted lines illustrate the location of shell-like structures. The red arrows mark the direction of the offset axis in the panels on the right.\label{fig:pvdout2}}
\end{figure}

We further investigate the regions 1 and 5 in position-velocity space. In the PVDs, we find a splitting of the CO (1-0) line intensity in the 500 pc ring NW and SE of the galaxy center. The slices through the data cube at regions 1 and 5 from Figure \ref{fig:dv} and the resulting PVDs are presented in Figure \ref{fig:pvdout2}. The line splitting at positions 1 and 5 (right panels on Figure \ref{fig:pvdout2}) extends over a region of \(\sim10\arcsec\) (500 pc) through the 500 pc ring, with morphology that can be explained as expanding molecular gas shells. Similar supershell (superbubble) structure has been observed in the 200 pc ring of M82 \citep{Wei99,Mat00,Mat05}, a nearby starburst galaxy which also exhibits a large-scale molecular outflow (e.g., \citealt{Nak87,Sal13}), as well as in NGC 2146 and NGC 3628 \citep{Tsa09,Tsa12}. A possible geometry of the supershells is illustrated in Figure \ref{fig:pvdout2} as black dashed lines.

The maximum splitting of the CO (1-0) line in the PVDs of Figure \ref{fig:pvdout2} is \(\Delta v_\mathrm{s}\sim150\) km s\(^{-1}\). Assuming spherical geometry (shell), this maximum velocity is related to the expansion velocity of the shells as \(v_\mathrm{exp}\approx \Delta v_\mathrm{s}/2\sim75\) km s\(^{-1}\). Although the uncertainty is large, this is within a factor of two comparable to the expansion velocity of the superbubble in M82 \citep{Wei99}.

The masses of the shells were estimated from the channel maps of the CO (1-0) data cube. Emission from the shells appears in channel maps in the velocity range of \(860~\mathrm{km~s}^{-1}<V_\mathrm{LSR}<1000~\mathrm{km~s}^{-1}\) (SE shell) and \(950~\mathrm{km~s}^{-1}<V_\mathrm{LSR}<1080~\mathrm{km~s}^{-1}\) (NW shell). Integrating the regions where the shell component appears yields fluxes of \(\mathcal{I}_\mathrm{CO}^\mathrm{SE}=2.663\) Jy km s\(^{-1}\) and \(\mathcal{I}_\mathrm{CO}^\mathrm{NW}=3.056\) Jy km s\(^{-1}\). The molecular gas masses can be calculated from Equation \ref{molmass}. Using \(X_\mathrm{CO}=0.8\times10^{20}~\mathrm{{cm}^{-2}(K~km~s^{-1})^{-1}}\), the derived molecular gas masses (corrected for the abundance of helium and other elements) are \(M_\mathrm{sh}^\mathrm{SE}\sim1.3\times10^6~M_\sun\) and \(M_\mathrm{sh}^\mathrm{NW}\sim1.5\times10^6~M_\sun\) for the SE and NW shell, respectively. The mass of the molecular gas shells is three orders of magnitude smaller than the total molecular gas mass in the starburst region (derived in section \ref{Db}). Including other outflow regions revealed in Figure \ref{fig:dv} and the nuclear outflow, we estimate that the total molecular gas mass in the wind does not exceed 10\% of the total molecular gas mass in the central 1 kpc, consistent with the results in \cite{Sal16}. For a shell size of \(r_\mathrm{sh}=300\) pc and expansion velocity \(v_\mathrm{exp}=75\) km s\(^{-1}\), the expansion time is \(\Delta t\sim r_\mathrm{sh}/v_\mathrm{exp}\sim4\times10^6\) yr, and the mass outflow rate is \(\Delta M_\mathrm{sh}/\Delta t\sim0.25~M_\sun~\mathrm{yr}^{-1}\), comparable to the SFR in regions R1-R5 discussed above. In the case of optically thin CO (1-0) emission, the mass outflow rate is an order of magnitude lower, implying inefficient suppression of star formation by the winds, i.e., a low mass loading factor \(\eta_\mathrm{m}\equiv(\Delta M_\mathrm{sh}/\Delta t)/\mathrm{SFR}\). Given the large gas surface density of \(\Sigma_\mathrm{mol}\sim300~M_\sun~\mathrm{pc}^{-2}\) in the 500 pc ring, the low \(\eta_\mathrm{m}<1\) is consistent with recent numerical simulations of supernova-driven winds that yield low values of \(\eta_\mathrm{m}\) for large \(\Sigma_\mathrm{gas}\) \citep{LBO17}.

The kinetic energy of a molecular gas shell is calculated from \(E_\mathrm{k}=M_\mathrm{sh}v_\mathrm{exp}^2/2\). Inserting the derived values above, the kinetic energies become \(E_\mathrm{k}^\mathrm{SE}\sim7\times10^{52}\) erg and \(E_\mathrm{k}^\mathrm{NW}\sim8\times10^{52}\) erg, respectively. In total, the energies are equivalent to \(\sim150\) Type II supernova explosions, each with a kinetic energy of \(E_0\sim10^{51}\) erg. From the rotational velocity of \(v_\mathrm{rot}=200\) km s\(^{-1}\) at the 500 pc ring, the orbital period of molecular gas is \(t_\mathrm{orb}=2\pi R/v_\mathrm{rot}\sim1.5\times10^7\) yr. If this period is the duration of the starburst during which supernovae exploded, the supernova rate in the 500 pc ring required to produce the molecular gas shells is of the order \(\mathcal{R}_\mathrm{SN}\sim\ E_\mathrm{sh}/E_0 t_\mathrm{orb}\sim10^{-5}\) yr\(^{-1}\). This result is two orders of magnitude lower than the estimated value of \(\mathcal{R}_\mathrm{SN}\) from radio and infrared observations in the ``hot spots'' in the entire central starburst region \citep{KSG94,Kot96}. Therefore, supernova explosions triggered by star formation activity in the 500 pc ring can be responsible for the expanding molecular gas shells. The spatial extent of the shells suggests that they have formed recently, within a few million years.

\section{Evolution of molecular clouds in the central 1 kpc starburst region}\label{E}

\subsection{GMC identification, distribution, and CO fluxes}\label{Ea}

In order to quantify the distribution and intensity of CO emission from giant molecular clouds (GMCs) in the central 1 kpc region, we carried out cloud identification using the CLUMPFIND algorithm (IDL version) \citep{WdB94}. The procedure was performed as follows. First, the algorithm was applied in 3 dimensions (position-position-velocity) to the CO (3-2) data cube (corrected for missing flux) to identify clumps. The CO (3-2) line was used instead of CO (1-0) because of higher angular resolution, where it was assumed that both lines trace the same bulk molecular gas (though the relative flux varies between 0.3 and 1; Figure \ref{fig:linrat}). The derived flux densities of clumps were then converted to those of CO (1-0) line by using the \(R_\mathrm{CO}\) ratio (section \ref{Ce}). Although the angular resolution of the \(R_\mathrm{CO}\) map is \(2\arcsec\), corresponding to the resolution of the CO (1-0) image, we converted the flux by taking average \(R_\mathrm{CO}\) values within regions of \(1\arcsec\) diameter toward the central positions of the clumps, corresponding to the major axis of the beam size of the CO (3-2) image. The distribution of the clouds and CO fluxes are analyzed for different regions defined by galactocentric rings (e.g., CND, CMC, 500 pc ring). The contribution from the outflow is not subtracted, but is expected to have a minor effect due to the low mass of the outflow gas relative to the total molecular gas mass in the central 1 kpc (section \ref{F}).

In the identification procedure, the minimum number of pixels needed to recognize a cloud (clump) was set to be the area of the beam, the beam size was 5.13 pixels, and the global r.m.s. noise level was set to 23 mJy beam\(^{-1}\) (about \(3~\sigma\) in emission-free channels). The program can identify but cannot resolve small molecular clouds (size \(\sim10\) pc). It is able to resolve GMCs and giant molecular associations (GMAs) larger than about 50 pc and perform deconvolution from the telescope beam.

The program identified a total of 214 clouds (clumps) within the imaged central starburst region, outputting the coordinates of their peaks and total CO (3-2) fluxes. The CO (3-2) flux of a clump is calculated as the sum of intensities in all \(i\)-pixels within the clump divided by the area of the beam (in pixels),

\begin{equation}
\left(\frac{S_\mathrm{CO(3-2)}}{\mathrm{Jy}}\right)=\left(\frac{\sum_i \mathcal{S}_i}{\mathrm{Jy~beam}^{-1}}\right)\left(\frac{A_\mathrm{b}}{\mathrm{beam^{-1}}}\right)^{-1},
\end{equation}
where \(A_\mathrm{b}=55.5\) beam\(^{-1}\). Multiplied by the velocity resolution element of the data cube, \(\Delta v=5\) km s\(^{-1}\), this gives the integrated flux of a clump, \(I_\mathrm{CO(3-2)}\) [Jy km s\(^{-1}\)], that, in principle, can be converted to the total molecular gas mass by using the \(R_\mathrm{CO}\) ratio map and a conversion factor \(X_\mathrm{CO}\) for the CO (1-0) line. Since low-\(J\) lines of \(^{12}\)CO are generally optically thick, the derived fluxes do not yield direct measurement of CO column densities.

\begin{figure}
\epsscale{1.2}
\plotone{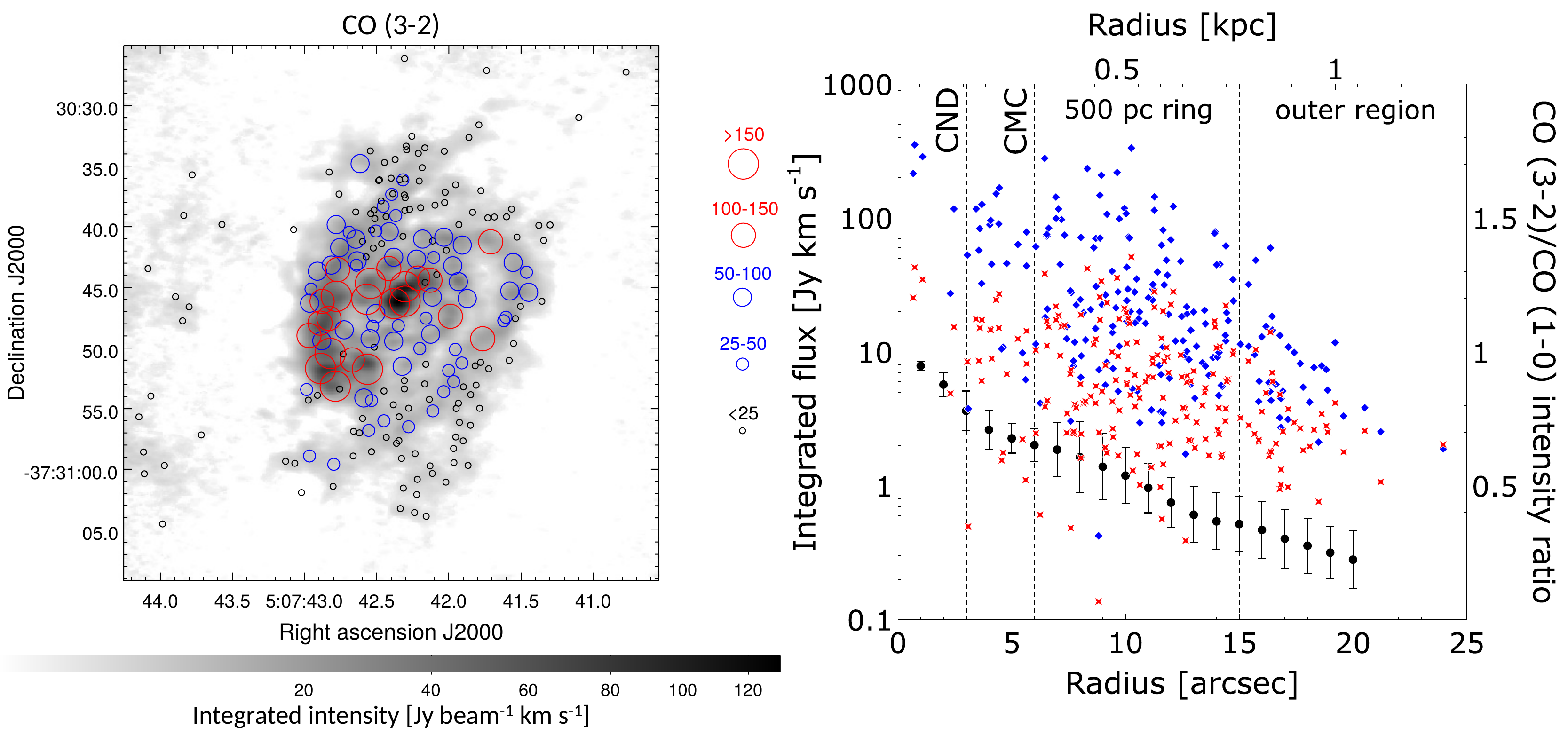}
\caption{Left) Distribution of identified clumps. The size of circles indicates the CO (3-2) integrated flux of each clump. Five flux-based types are plotted to guide the eye. The CO (3-2) intensity is presented on greyscale. Right) CO (1-0) flux (red stars) and CO (3-2) flux (blue diamonds) of each clump plotted as a function of distance from the galactic center (deprojected, i.e., corrected for galactic inclination and position angle). Also shown is the azimuthally-averaged CO (3-2)/CO (1-0) intensity ratio (black circles; right vertical axis).\label{fig:clumps1}}
\end{figure}

The objects are identified within galactocentric rings as follows: 5 clumps in the CND (\(r<3\arcsec\)), 21 in the CMC region (\(3\arcsec<r<6\arcsec\)), 135 in the 500 pc ring (\(6\arcsec<r<15\arcsec\)), and 52 beyond the ring (\(r>15\arcsec\)). Figure \ref{fig:clumps1} shows the clump distribution across the central starburst region and CO integrated flux as a function of radius. The right panel in the figure shows both CO (1-0) and CO (3-2) fluxes and the azimuthally-averaged CO (3-2)/CO (1-0) intensity ratio (\(R_\mathrm{CO}\)). Note that clouds with larger CO (3-2) fluxes are typically detected closer to the galactic center. All clouds beyond the 500 pc ring have integrated fluxes \(I_\mathrm{CO(3-2)}\lesssim25\) Jy km s\(^{-1}\). Although the scatter is large, there is a tendency of a decrease in cloud fluxes with galactocentric radius. Figure \ref{fig:clumps1} shows that: (1) the clouds in the CND are typically luminous, (2) the fraction of clouds with intermediate fluxes (\(50~\mathrm{Jy~km~s^{-1}}<I_\mathrm{CO(3-2)}<100\) Jy km s\(^{-1}\)) is larger in the CMC region than in the 500 pc ring, and (3) only low-luminosity clouds are present beyond the 500 pc ring. Although the number of clouds is largest in the 500 pc ring, 101 (75\%) of them have fluxes \(I_\mathrm{CO(3-2)}<50\) Jy km s\(^{-1}\). Most clouds have integrated fluxes of \(I_\mathrm{CO(1-0)}\sim5\) Jy km s\(^{-1}\), corresponding to a mass of \(2\times10^6~M_\sun\) for \(X_\mathrm{CO}=0.8\times10^{20}~\mathrm{{cm}^{-2}(K~km~s^{-1})^{-1}}\). These objects lie in the range of giant molecular clouds (GMCs).\footnote{For example, \cite{SSS85} define clouds with \(M>10^5~M_\sun\) as GMCs.} Clouds with fluxes above \(I_\mathrm{CO(1-0)}\gtrsim20\) Jy km s\(^{-1}\) can be regarded as giant molecular associations (GMAs) with masses \(\sim10^7~M_\sun\). The largest number of GMAs (\(\sim10\)) is found in the 500 pc ring.

\subsection{GMC evolution: inflow, star formation, and outflow}\label{Eb}

\begin{figure}
\epsscale{1}
\plotone{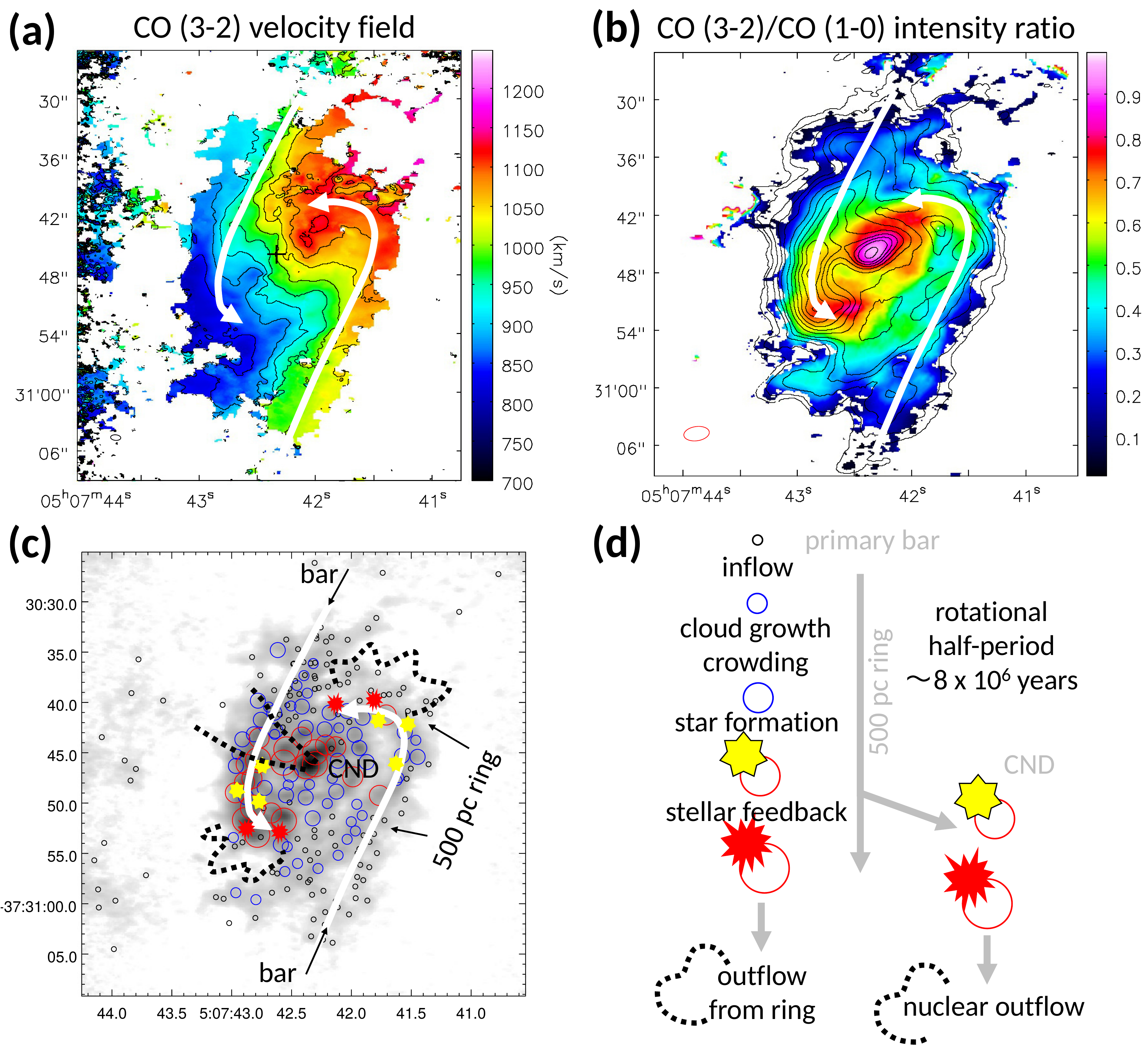}
\caption{Illustration of the evolution of molecular gas in the central starburst 1-kpc region. (a) Velocity field (moment 1) of CO (3-2) data. The contours are plotted from 800 km s\(^{-1}\) to 1200 km s\(^{-1}\) in steps of 40 km s\(^{-1}\). The black plus symbol marks the galactic center. (b) CO (3-2) to CO (1-0) intensity ratio. (c) GMC luminosities and star formation activity. (d) Evolution of molecular clouds from bar-induced inflow to outflow feedback. The white arrows indicate the inflow and rotational motion of molecular gas.\label{fig:evo}}
\end{figure}

The distribution of clumps presented in the previous section can be explained in terms of GMC evolution. Gas dynamics is one of the key factors that determine the large-scale motion of GMCs in galaxies. Observations and numerical simulations show that gas motions in a bar potential are non-circular and lead to a gradual inflow (e.g., \citealt{WH95,SBS00,She05,Com14}). In galaxies with inner Lindblad resonances, gas can accumulate on approximately circular orbits near the galactic center (e.g., \citealt{But86,Ath92,Mac02}), and this is the case in NGC 1808 where an inner resonance has been found close to the radius of the 500 pc ring \citep{Sal16}. In the case of NGC 1808, large non-circular motions were discovered in the kpc-scale disk of the galaxy that can be attributed to the primary bar. The non-circular motions can be seen in Figure \ref{fig:evo} (panel a) as the overall ``S'' shape of the velocity field, as well as in Figure \ref{fig:dv}(c); the distortions from circular rotation are most prominent beyond the 500 pc ring, in the regions N and S of the galactic center. Molecular clouds in the bar are not on circular orbits, but lose angular momentum to the bar and slowly fall toward lower orbits eventually reaching the 500 pc ring. The large-scale inflow motion is illustrated with white arrows in Figure \ref{fig:evo} (see  \cite{Kor96} and section 4.2 of \cite{Sal16} for discussions about non-circular motions of gas in the bar of NGC 1808).

As shown above, the orbital period of molecular gas in the 500 pc ring is of the order of \(t_\mathrm{orb}=1.5\times10^7\) years. Figure \ref{fig:clumps1} shows that the luminosity function of GMCs on the orbit from the primary bar toward the 500 pc ring is not constant. In particular, the CO fluxes are found to increase in the rotational direction (indicated by white arrows in Figure \ref{fig:evo}) over the rotational half-period, traveling approximately one half of the circumference of the 500 pc ring upon entering the ring from the leading ridge of the primary bar. Based on the distribution and fluxes of GMCs, we propose the following dynamically-driven evolutionary sequence of the GMCs in the central starburst region, as illustrated in Figure \ref{fig:evo}(d):

\begin{enumerate}

\item{Low CO luminosity clouds (integrated flux \(I_\mathrm{CO(3-2)}<25\) Jy km s\(^{-1}\) or \(I_\mathrm{CO(1-0)}<10\) Jy km s\(^{-1}\)) enter the 500 pc ring from the large-scale bar, driven by the gravitational torque from the non-axisymmetric potential of the bar. This is the phase of cloud accretion onto the 500 pc ring. Non-detection of radio continuum and low \(R_\mathrm{CO}\) ratio indicate low SFR in the accreting material.}

\item{Upon accreting onto the 500 pc ring, the clouds grow via collisions with each other and with clouds that are already in circular motion in the ring. Gravitational collapse in low-shear conditions and/or cloud-cloud collisions trigger star formation, observationally identified as an enhanced \(R_\mathrm{CO}\) ratio and continuum emission at 93 GHz and lower frequencies (section \ref{Ce}).}

\item{Inside the ring, star formation produces massive young stars that generate stellar winds and explode as supernovae on a timescale of several million years, comparable to one quarter -- one half of the travel time (orbital period) of the clouds in the ring. These regions are seen as the ``hot spots'' detected at various wavelengths.}

\item{Feedback from massive stars drives molecular outflows (supershells) that eject molecular gas from the ring (section \ref{Dc}). At the same time, the motion of molecular gas in the ring is affected by the gravitational potential of the nuclear bar. The ``bars-within-bars'' effect \citep{SFB89} drives molecular gas from the ring toward the CND. The inflowing gas is distributed in a spiral pattern and forms a star-forming ring with CMCs at radius \(r\sim200\) pc (section \ref{Ca}).}

\item{Accumulation of molecular gas in the CND and around it (in CMCs) fuels an intense nuclear starburst. Its feedback is the origin of the nuclear molecular gas outflow discussed in section \ref{Dc0}. The gas reservoir in the CND is further replenished by inflow.}

\end{enumerate}

This evolutionary scenario can explain not only the distribution of GMCs, but also the base of molecular gas outflows in the 500 pc ring and the enhancement of \(R_\mathrm{CO}\) in the CND, CMCs, and the downstream regions of the 500 pc ring (Figure \ref{fig:linrat}). However, we emphasize that since the analysis of the conversion factor \(X_\mathrm{CO}\) is beyond the scope of this paper, the proposed evolutionary sequence should be regarded in the aspect of CO luminosity and excitation rather than molecular cloud mass. Further study is necessary to probe the variation of \(X_\mathrm{CO}\) (cloud mass function) across the starburst region from the CND to beyond the 500 pc ring. Also, multi-line observations of other starburst galaxies at high resolution and sensitivity will reveal whether the evolutionary sequence of molecular clouds is typical in barred starburst galaxies with central rings, as observed in NGC 1808.

\section{Summary}\label{G}

We have presented new CO and continuum observations of the central 1 kpc of the nearby starburst galaxy NGC 1808 carried out with ALMA and VLA. Below is a summary of the main findings:

\begin{enumerate}

\item{Continuum emission at 21, 32, 93, and 350 GHz was detected toward the central region at high angular resolution. The continuum is distributed in a number of discrete sources that coincide with the radio ``hot spots'' detected earlier at lower frequencies forming a ring of radius \(\sim200\) pc. The derived spectral energy distribution of the core indicates prominent free-free emission at 93 GHz arising from star-forming regions.}

\item{The distribution of the CO (3-2) intensity in the central 1 kpc confirms the presence of a circumnuclear disk (CND) in the central 100 pc, nuclear (molecular) spiral arms, 500 pc pseudoring, and non-circular motions, that were previously observed in high-resolution CO (1-0) observations. The region between the CND and the ring is resolved to a number of compact sources referred to as the central molecular clouds (CMCs), that coincide with the continuum ``hot spots''.}

\item{The distribution of CO (3-2) reveals a double peak in the CND, interpreted as a molecular gas torus with a radius of \(\sim30\) pc. The core, luminous in continuum and CO, is located between the peaks. While the torus exhibits rotational motion consistent with the overall ``rigid-body'' rotation in the central region, the core shows a peculiar velocity component, that indicates the presence of molecular gas in streaming motion or fast rotation around the galactic center.}

\item{Using CO (1-0) and 350 GHz continuum as tracers of gas mass, and 93 GHz as a tracer of SFR, we derived the SFE in the CND and CMCs at the resolution of 100 pc. The SFE is of the order \(\sim10^{-8}\) yr\(^{-1}\) in these regions, implying a depletion time an order of magnitude shorter than the typical value in the disks of nearby galaxies, consistent with the starburst regime of the Kennicutt-Schmidt law.}

\item{The presence of a nuclear gas stream, with a non-circular motion amplitude as large as \(\gtrsim100\) km s\(^{-1}\), interpreted as an outflow, is confirmed in the new CO (3-2) data. A velocity gradient along the outflow is measured to be \(\Delta v/\Delta h\sim+0.4\) km s\(^{-1}\) pc\(^{-1}\), indicating acceleration to a height of \(h\sim900\) pc. The line intensity ratio of CO (3-2) to CO (1-0) in the nuclear outflow is decreasing from 0.5 to 0.2 along the outflow, indicating a change in physical conditions.}

\item{In addition to the nuclear wind, we report the detection of supershell outflows from the 500 pc ring. The mass of the outflows is \(\sim10^{-3}\) of the total mass in the starburst region, and the mass outflow rate is \(\sim10\%\) of the star formation rate in the case of optically thin CO (1-0) emission. This suggests that star formation in the 500 pc ring is not significantly suppressed by the molecular winds at this stage.}

\item{The clump identification program CLUMPFIND was applied to make a quantitative analysis of the distribution and CO fluxes of molecular clouds in the central 1 kpc starburst reigon. The CO fluxes of the cloud populations vary with galactocentric radius in the sense that clouds tend to have higher CO (3-2)/(1-0) intensity ratio in regions of active star formation. The distribution of clouds and the CO (3-2) to CO (1-0) line intensity ratio strongly suggests that clouds evolve as they accrete onto the 500 pc ring from the large-scale bar, fuel star formation, and undergo ejection in the form of outflows as star formation feedback.}

\end{enumerate}

\acknowledgements

The authors thank the anonymous referee for many useful comments that improved the structure of the paper. D.S. is grateful to professor M. Seta for support in data reduction. This paper makes use of the following ALMA data: ADS/JAO.ALMA\#2012.1.01004.S and \#2013.1.00911.S. ALMA is a partnership of ESO (representing its member states), NSF (USA), and NINS (Japan), together with NRC (Canada) and NSC and ASIAA (Taiwan), in cooperation with the Republic of Chile. The Joint ALMA Observatory is operated by ESO, AUI/NRAO, and NAOJ. The National Radio Astronomy Observatory is a facility of the National Science Foundation operated under cooperative agreement by Associated Universities, Inc. Based on observations made with the NASA/ESA \emph{Hubble Space Telescope} and obtained from the Hubble Legacy Archive, which is a collaboration between the Space Telescope Science Institute (STScI/NASA), the Space Telescope European Coordinating Facility (ST-ECF/ESA), and the Canadian Astronomy Data Centre (CADC/NRC/CSA). This research has made use of the NASA/IPAC Extragalactic Database (NED), which is operated by the Jet Propulsion Laboratory, California Institute of Technology, under contract with the National Aeronautics and Space Administration. D.S. was supported by the ALMA Japan Research Grant of NAOJ Chile Observatory, NAOJ-ALMA-59 and NAOJ-ALMA-98.

\end{document}